\documentstyle[prb,aps,multicol,epsf]{revtex}
\title{Electronic states, Mott localization, 
	electron-lattice coupling, and dimerization 
	for correlated one-dimensional systems.\ II.}
\author{Adam Rycerz and Jozef Spa{\l}ek, \\
	Marian Smoluchowski Institute of Physics,
	Jagiellonian University, \\
	ulica Reymonta 4, 30-059 Krak\'{o}w, Poland}
\def\date{}

\newcommand{\propar}{\par}

\newcounter{Romenum}

\newcounter{romenum}
\renewcommand{\theromenum}{\roman{romenum}}

\newcommand{\simgeq}%
{\raisebox{-.5ex}{$\stackrel{\textstyle >}{\textstyle\sim}$}}
\newcommand{\simleq}%
{\raisebox{-.5ex}{$\stackrel{\textstyle <}{\textstyle\sim}$}}
\newcommand{\eqau}{\stackrel{\rm a.u.}{=}}
\newcommand{\bra}[1]{\left<#1\right|}
\newcommand{\ket}[1]{\left|#1\right>}
\newcommand{\newton}[2]%
{\left(\raisebox{-1ex}{$\stackrel{\textstyle #1}{#2}$}\right)}
\newcommand{\half}%
{\frac{\scriptstyle 1}{\scriptstyle 2}}

\begin{document}
\maketitle

\begin{abstract}
We discuss physical properties of strongly correlated electron states
for a linear chain obtained with the help of 
the recently proposed new method combining the exact
diagonalization in the Fock space with an {\em ab initio} readjustment 
of the single-particle orbitals in the correlated state. 
The method extends the current discussion of the correlated 
states since the properties are obtained with varying lattice spacing. 
The finite system of $N$ atoms {\em evolves} with the increasing
interatomic distance from a Fermi-liquid-like state into the Mott insulator.
The criteria of the localization are discussed in detail
since the results are already convergent for $N\geq 8$.
During this process the Fermi-Dirac distribution gets smeared out, 
the effective band mass increases by $\sim 50\%$, 
and the spin-spin correlation 
functions reduce to those for the Heisenberg antiferromagnet. 
Values of the microscopic parameters such as the hopping 
and the kinetic-exchange integrals, as well as the magnitude of both
intra- and inter-atomic Coulomb and exchange interactions are calculated. 
We also determine the values of various local electron-lattice couplings 
and show that they are {\em comparable} to the kinetic exchange contribution 
in the strong-correlation limit. 
The magnitudes of the dimerization and the zero-point motion are also 
discussed. Our results provide a canonical 
example of a tractable strongly correlated system with a precise,
first-principle description as a function of interatomic distance
of a model system involving {\em all} hopping integrals,
{\em all} pair-site interactions, and the {\em exact} one-band 
Wannier functions. \\
\vspace{0in} 
\end{abstract}

\begin{multicols}{2}
\narrowtext

\section{Introduction}
\label{intro}

In spite of the enormous successes of the approach based on the
effective single-particle wave equation \cite{jogu} for many 
$3$-dimensional metals
and semiconductors, the understanding of the so-called {\em correlated 
fermionic systems} is still lacking, particularly for the systems
of lower dimensionality, $d=1$ and $2$. This is because in their 
description of the electronic states 
the role of the long-range Coulomb interaction is crucial, as the charge 
screening becomes less effective \cite{hubb}. In result, the interaction
cannot be regarded as small and phenomena
such as the spin-charge separation \cite{schu}, the Mott metal-insulator
transition \cite{poib} or
the strong electron-lattice interaction (leading among others to the 
Peierls distorted \cite{peie} state) 
appear in various quasi-one-dimensional systems. In addition,
a normal metallic state in quantum wires \cite{goni} and nanotubes 
\cite{mint}, as well as the superconducting state in organic $1d$ metals 
are also observed,
raising the question about the relative role of the single-particle
dynamics and the repulsive Coulomb interactions treated in 
a nonperturbational manner. 
To phrase it differently, in one-dimensional systems we
consider here the single-particle states and
the Coulomb interactions are treated on equal footing from 
the outset.

We have recently proposed \cite{spapo,ryspa} (those References are 
regarded as Part I) a new approach to the 
correlated fermion systems, which provides rigorous results for
model (one-band) systems containing $N\leq 12$ atoms.
The results concerning the principal local characteristics are very
rapidly converging and provide a reliable estimate of
the system characteristics such as the microscopic
interaction parameters, the ground state energy, the magnitude
of Peierls distorsion or the zero point motion, all as function
of the interatomic distance. It is the purpose of this
paper to provide a complete picture of the 
electronic and lattice properties by providing the band effective 
masses, the magnitude of exchange interactions and related to them
the spin-spin correlation functions,  
the detailed discussion of a gradual transition from a metal to the
Mott insulator, and the constants of the local
coupling of electrons to the lattice. 
We also consider the dimerized state of the system. We believe that
analysis of this type containing 
a combination of a rigorous treatment of the interactions 
in the Fock space combined with {\em ab initio} treatment 
of the single-particle wave-functions 
in the Hilbert state, that are allowed to relax in the
correlated state, provides the first step in a rigorous quantitive 
analysis of linear-chain of finite size and {\em correlated} quantum dot 
systems. In other words, our first-principle method is specifically designed
to treat such correlated systems, although some general conclusions are
also discussed.

The structure of the paper is as follows. In the next section
(and in Appendix~\ref{micro}) we summarize our
method and provide the details not published before. 
In Section~\ref{meta} the crossover from metallic to the Mott 
insulating state is dealt with to provide the proper ground state for
a subsequent analysis. 
In Section~\ref{kin} we discuss the magnetic (kinetic-exchange)
interactions and the spin-spin correlation functions. 
In Section~\ref{elec} we determine the local electron-phonon coupling
constants and compare their magnitude with that of the magnetic
interactions. The main feature of the above
analysis is to provide the properties as a function of 
the lattice spacing. In this respect our approach differs from the 
numerous solutions \cite{hubb}-\cite{poib} of parametrized models 
such as the (extended) Hubbard model, where the physical properties are 
discussed as a function of the model parameters. 
Here {\em all} one- and two-site parameters are calculated explicitly 
and this feature 
allows for a consideration of e.g.\ one-band system with inclusion of 
both the {\em exact} Wannier functions, {\em all} hopping integrals and 
{\em all} two-site interactions.

\section{Method}
\label{meto}

We start with the Hamiltonian containing $N$ lattice sites
with {\em all} hopping integrals $t_{ij}$ and with {\em all} 
two-site interactions, 
which for the linear chain with periodic boundary conditions can be
written as
$$
  {\cal H}=\sum_{i=0}^{N-1}\left\{\epsilon_an_i+
  Un_{i\uparrow}n_{i\downarrow}+
  \sum_{j=0}^{i-1}\left[
  \left(K_{ij}-\frac{1}{2}J_{ij}\right)n_in_j \right.\right.
$$
$$
  -2J_{ij}{\bf S}_i\cdot{\bf S}_j
  +\sum_{\sigma}\left(t_{ij}+V_{ij}(n_{i\bar{\sigma}}+
  n_{j\bar{\sigma}})\right)
  \left(a^{\dagger}_{i\sigma}a_{j\sigma}+{\rm h.c.}\right)
$$
\begin{equation}
 \label{ham:lanext}
  \left.\left.
  +J_{ij}\left(a^{\dagger}_{i\uparrow}a^{\dagger}_{i\downarrow}
  a_{j\downarrow}a_{j\uparrow}+{\rm h.c.}\right)\right]\right\}.    
\end{equation}

The first term represents the atomic energy (we include it 
explicitly, since $\epsilon_a$ changes with the varying lattice constant).
The second describes the intraatomic Coulomb interaction 
(the Hubbard term). 
The next two terms represent the direct intersite Coulomb 
interaction ($\sim K_{ij}$) and the Heisenberg-Dirac exchange 
($\sim J_{ij}$). 
The fifth and the sixth term express respectively the single-particle 
($\sim t_{ij}$) and the correlated-hopping ($\sim V_{ij}$) terms,
whereas the last includes the pair-electron hopping 
$i\rightleftharpoons j$. 
The microscopic parameters have been defined before 
(cf.\ Appendix~\ref{single} of Ref.~\cite{spapo}).
Their values as a function of the interatomic distance $R$
have been determined before \cite{ryspa}; here
we provide in Table~\ref{tab:ukvj} the values of the interaction
parameters together with their asymptotic analytic expressions, 
which reproduce well their values for 
$R\simgeq 4a_0$ ($\simgeq 2\AA$).

One should underline that the parameters are here 
defined in terms of exact Wannier function 
for this one-band (one-orbital-per-site) system, that
is defined as 
\begin{equation}
\label{wann}
  w_i({\bf r})=\sum_{j=1}^N\beta_{i-j}\Psi_j({\bf r}),
\end{equation}
where $\Psi_j({\bf r})$ represents atomic function of $s$-type 
centered on site $j$:
\begin{equation}
\label{atfun}
  \Psi_j({\bf r})\equiv\Psi\left({\bf r}-{\bf R}_j\right)
  =\left(\frac{\alpha^3}{\pi}\right)^{1/2}
  \exp\left(-\alpha\left|{\bf r}-{\bf R}_j\right|\right). 
\end{equation}
The quantity $\alpha$, which represents the inverse orbital size
(and calculated in units of the Bohr radius $a_0$) will
be determined by optimizing the orthonormal atomic (Wannier)
basis in the correlated state. In effect, the parameters
are $\epsilon_a=\left<w_i\right|T\left|w_i\right>$,
$t_{i-j}=\left<w_i\right|T\left|w_j\right>$,
$U=\left<w_iw_i\right|V_{12}\left|w_iw_i\right>$,
$K_{i-j}=\left<w_iw_j\right|V_{12}\left|w_iw_j\right>$,  
$J_{i-j}=\left<w_iw_j\right|V_{12}\left|w_jw_i\right>=
\left<w_iw_i\right|V_{12}\left|w_jw_j\right>$, and 
$V_{i-j}=\left<w_iw_i\right|V_{12}\left|w_iw_j\right>$. 
The operator $T$ represents the full single-particle lattice potential, i.e.
$$
  T({\bf r})=-\frac{\hbar^2}{2m}\nabla^2
  -\sum_{j=0}^{N-1}\frac{e^2}{\left|{\bf r}-{\bf R}_j\right|}
$$
\begin{equation}
\label{oper:hopp}
  \eqau-\frac{1}{2}\nabla^2
  -\sum_{j=0}^{N-1}\frac{2}{\left|{\bf r}-{\bf R}_j\right|},
\end{equation}
where $\mbox{a.u.}$ means the expression 
in  atomic units. $V_{12}=e^2/|{\bf r}_1-{\bf r}_2|$
is the usual Coulomb potential
(we do not include any screening by e.g.\ core 
electrons as we want to discuss the model situation, 
but in a rigorous manner). 

\begin{table}
\caption{
Microscopic parameters values for various interaction
terms in the starting Hamiltonian for $N=8$ sites.
\label{tab:ukvj}}
\begin{tabular}{ r|r|r|r|r }
$R$ [$a_0$] & $U$ [Ry] & $K_1$ [Ry] & $K_2$ [Ry] & $K_3$ [Ry] \\ \tableline
2.0 & 2.301 & 1.077 & 0.676 & 0.450 \\ 
2.5 & 1.949 & 0.843 & 0.499 & 0.331 \\ 
3.0 & 1.717 & 0.692 & 0.391 & 0.259 \\ 
4.0 & 1.452 & 0.508 & 0.269 & 0.179 \\ 
5.0 & 1.327 & 0.403 & 0.206 & 0.138  \\ 
$R\rightarrow\infty$ & 1.250 & $2/R$ & $1/R$ & $2/3R$ \\ 
\end{tabular}
\begin{tabular}{ r|r|r|r }
$R$ [$a_0$] & $J_1$ [mRy] & $V_1$ [mRy] & $V_2$ [mRy] \\ \tableline
2.0 & 9.54 & -18.07 & 33.58 \\ 
2.5 & 7.39 & -17.45 & 19.58 \\ 
3.0 & 5.59 & -16.08 & 11.95 \\ 
4.0 & 2.90 & -12.92 & 4.49 \\ 
5.0 & 1.26 & -9.64  & 1.56 \\ 
$R\rightarrow\infty$ & 
$\frac{4R^3}{15}e^{-R}\log R$ & $2Re^{-R}$ & $4Re^{-2R}$ \\ 
\end{tabular}
\end{table}

\begin{figure}
\centering\epsfxsize=8cm\epsffile{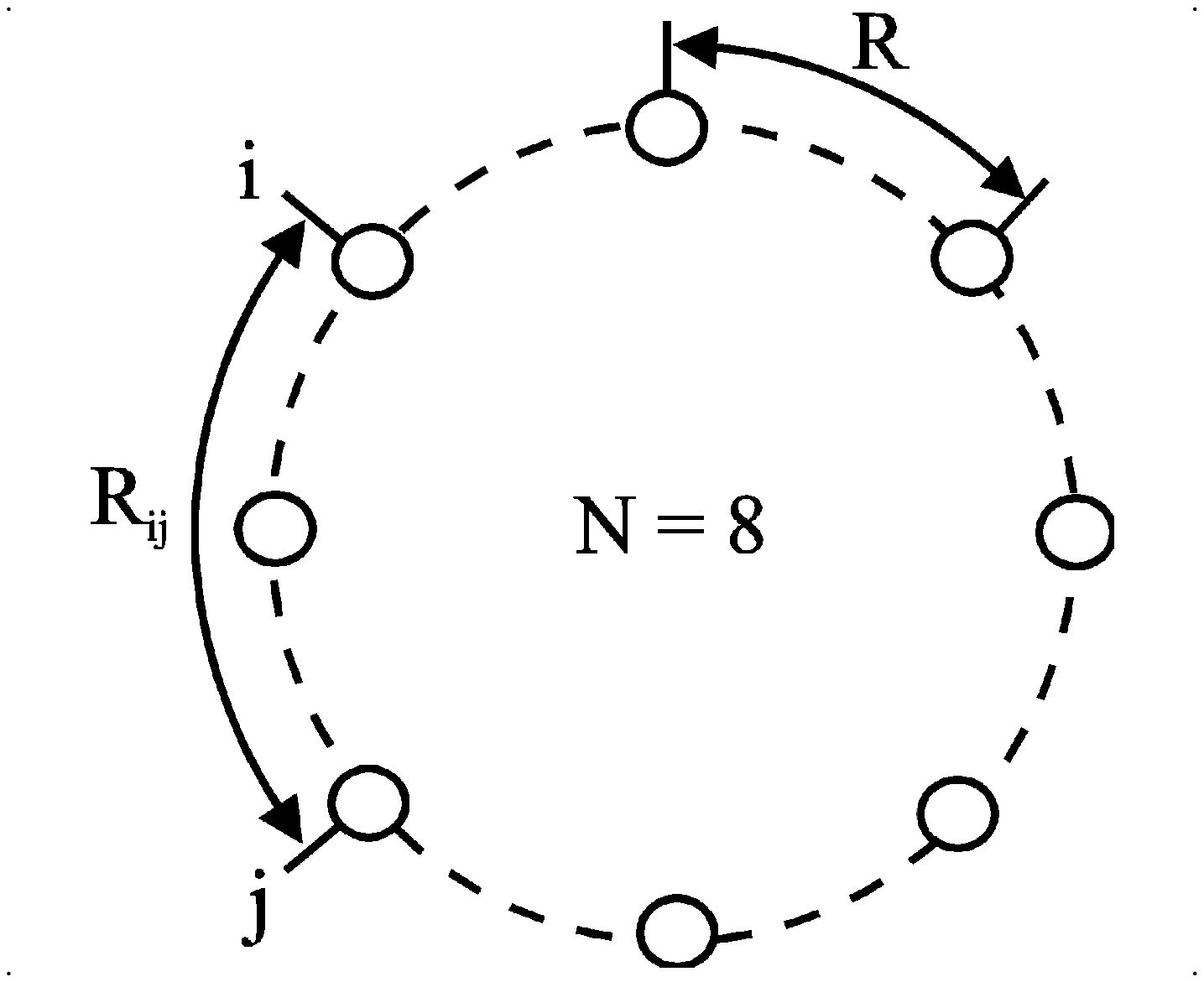} \par
\vspace{0.5cm}
\caption{
Schematic representation of the finite linear chain with
periodic boundary conditions used in the calculations.
\label{rys:lanext}}
\end{figure}

The inclusion of the full Coulomb potential (\ref{oper:hopp})
will introduce three-site terms already in the hopping integral
$t_{ij}'$ in the atomic basis; this feature follows the
definitions and expressions for {\em all} the interaction parameters 
for the Wannier function (\ref{wann}) provided in Appendix~\ref{micro} 
(Ref.~\cite{spapo} does not contain 
all parameters explicitly and the evaluation of $t_{ij}$ is absent, 
see Appendix~\ref{single}). Also, the outlines of the diagonalization
method and of the single-particle-basis optimization are briefly
summarized in Appendix~\ref{ground}. 
In what follows we concentrate on the physical properties of the
results for a linear chain with periodic boundary conditions 
containing up to $N=12$ atoms. 
Particular emphasis is put on the asymptotic properties
(i.e. those weakly dependent on $N$) of those truly nanoscopic 
systems depicted schematically in Fig.~\ref{rys:lanext}.

\section{Metallic and Mott localized states}
\label{meta}

\subsection{Band energy and the effective band mass}

\begin{figure}
\centering\epsfxsize=8cm\epsffile{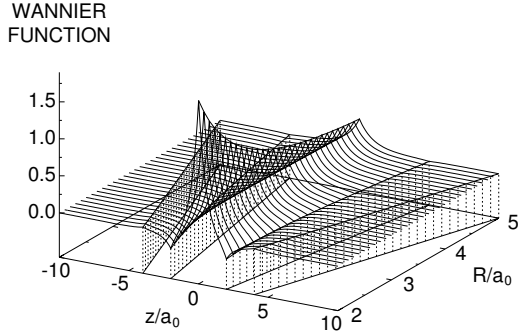} \par
\caption{
Space profile of the Wannier-function evolution with 
the increasing interatomic distance.
\label{rys:wann}}
\end{figure}

We begin with the whole analysis of the band electron
properties. The values of atomic level position 
$\epsilon_a=\int d^3{\bf r}w_i^*({\bf r})T({\bf r})w_i({\bf r})$ 
and of the hopping matrix elements 
$t_m=\int d^3{\bf r}w_i^*({\bf r})T({\bf r})w_{i+m}({\bf r})$ 
with $m=1,\dots,5$ are listed in Table~\ref{tab:hopp} for 
$R/a_0=2\div 5$. 
At most, three first hopping matrix elements are important;
this is sufficient to achieve the asymptotic properties
already for $N\geq 8$. Such a fast diminution of $t_m$ with
increasing $m$ is due to the reduction of the spread of the Wannier 
functions $\{w_i({\bf r})\}$ due to the intersite repulsive Coulomb
interaction \cite{ryspa}. The space profile of the Wannier 
function along the chain direction for different interatomic distances 
is shown in Fig.~\ref{rys:wann}. 
The wave-function renormalization is apparent 
for smaller $R/a_0$, where metallic state is expected to appear. 

From Table~\ref{tab:hopp} one can also see that the atomic energy is
strongly dependent on the distance $R$ and even for $R=5a_0$ 
it is decisively lower ($\approx-2.7\mbox{ Ry}$) than that in the $1s$ 
state of hydrogen atom ($-1\mbox{ Ry}$). This decrease is caused both 
by the long-range nature of the attractive potential $V({\bf r})$. 

\begin{figure}
\centering\epsfxsize=8cm\epsffile{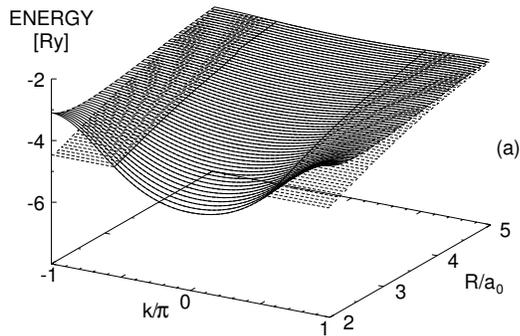} \propar
\centering\epsfxsize=8cm\epsffile{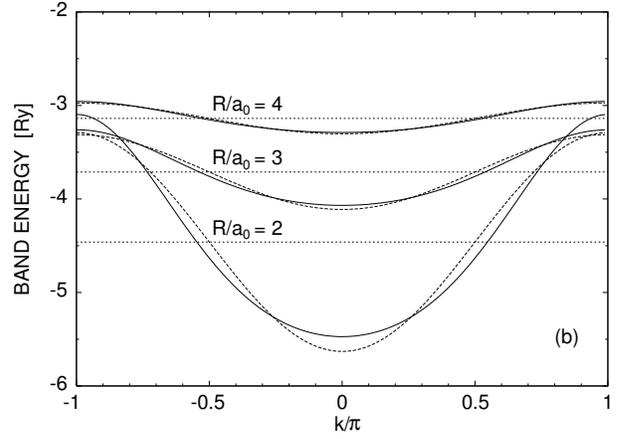} \par
\vspace{0.5cm}
\caption{
a) The space profile of the band shape for $N=10$ versus $R$, 
taking into account the calculated hopping integrals 
$\{t_m\}_{m=1,\dots,5}$. 
The horizontal plane intersecting the band marks the Fermi level
position; b) the flattening of the band shape with the increasing
$R$ (the dashed lines describe the band 
shape if only $t_1$ is included).
\label{rys:eband}}
\end{figure}

The band energy can be calculated directly by introducing 
the corresponding single-particle energy
\begin{equation}
\label{eband}
  \epsilon_n=\sum_{m=0}^{N-1}t_m\cos\left(\frac{2\pi mk_n}{N}\right),
\end{equation}
where $k_n=-[N/2],\dots,[(N-1)/2]$ ($[x]$ denotes the integer part
of $x$) is the quantum number (related to the wavevector via relation 
$k=2\pi k_n/NR$) in the first Brillouin zone.
The profile of the bands evolving with the increasing 
lattice parameter $R$ is shown in Fig.~\ref{rys:eband}ab. 
Only for the distances 
$R\simgeq 4a_0$ we can approximate the band with one hopping 
parameter $t_1$, when the bandwidth is strongly reduced, 
as one would expect on the basis of the tight-binding-approximation
(TBA). The Fermi level is always 
at the point $kR/\pi=0.5$, as we have assumed that we have 
one electron per atom. 

\begin{figure}
\centering\epsfxsize=8cm\epsffile{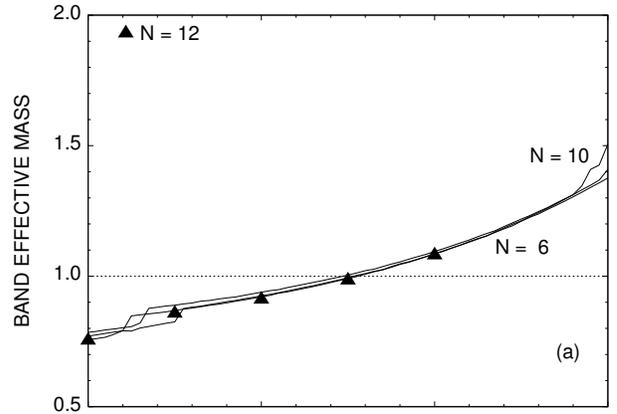} \propar
\centering\epsfxsize=8cm\epsffile{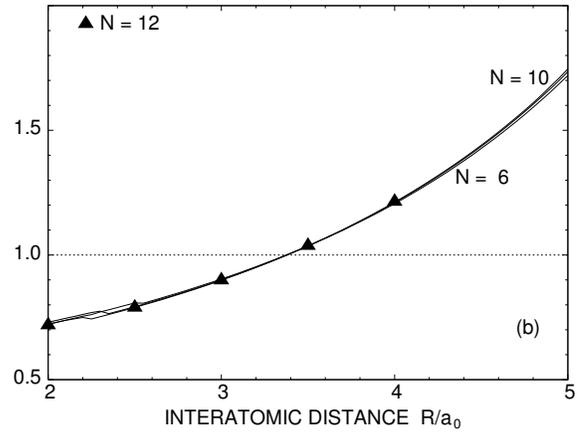} \par
\vspace{0.5cm}
\caption{
a) The band effective mass at the band center ($k=0$)
and b) for the Fermi wave vector $k=k_F$, both vs.\ $R$.
Note a general insensitivity of the results for different $N$.
\label{rys:meff}}
\end{figure}

\begin{figure}
\centering\epsfxsize=8cm\epsffile{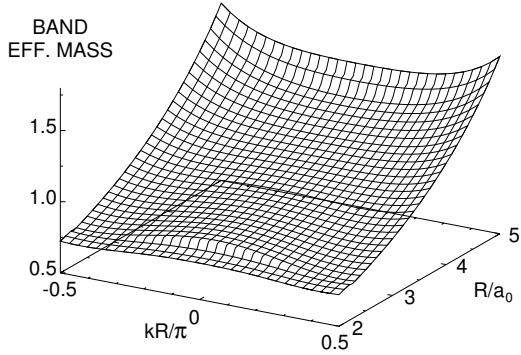} \par
\caption{
The profile of the (bare)
band effective mass as a function of the wave vector $k$
and the interatomic distance $R$ (for $N=10$).
\label{rys:meffsurf}}
\end{figure}

From the band energies we can obtain the effective 
band mass at either the band center ($k=0$) or at the 
Fermi point ($k_F=\pi/2R$):
\begin{equation}
\label{meff:zero}
  m_0^*={\hbar}^2
  \left(\left.\frac{d^2\epsilon_k}{dk^2}\right|_{k=0}\right)^{-1}
  \eqau\frac{2}{R^2}
  \left(\left.\frac{d^2\epsilon_k}{dk^2}\right|_{k=0}\right)^{-1},
\end{equation}
\begin{equation}
\label{meff:fermi}
  m_F^*={\hbar}^2k_F
  \left(\left.\frac{d\epsilon_k}{dk}\right|_{k=k_F}\right)^{-1}
  \eqau\frac{\pi}{R^2}
  \left(\left.\frac{d\epsilon_k}{dk}\right|_{k=\pi/2}\right)^{-1}.
\end{equation}
\noindent
The relative effective mass $m^*/m_e$ (where $m_e$ is the free
rest mass) are plotted in Fig.~\ref{rys:meff}ab (the inaccuracies are
due to the numerical differentiation of $\epsilon_k$). The mass
grows with the increasing interatomic distance and
reaches about $50-70\%$ higher value (then $m_e$) for $R/a_0\approx 5$. 
One should also note that even thought the Wannier 
functions are obtained from optimizing the energy of the
interacting state, the masses are light ($m^*<m_e$) for $R\leq 3.5a_0$.
What is more important, they practically do not depend on $N$
for $N\geq 6$. 
Therefore, in Fig.~\ref{rys:meffsurf} we have plotted the profile
of $m^*/m_e$ only for $N=10$ versus both $R/a_0$ and $kR/\pi$.

The calculated band energies will serve as an input 
in the discussion of the relative role of the Coulomb interaction at
the onset of the Mott localized state. This is
considered next by determining first the bare bandwidth
$W=2\left|\sum_{m=1}^{N-1}t_m\right|$ and comparing it with the 
magnitude of the effective Coulomb interaction.

\subsection{Onset of the Mott localized state}

\begin{figure}
\centering\epsfxsize=8cm\epsffile{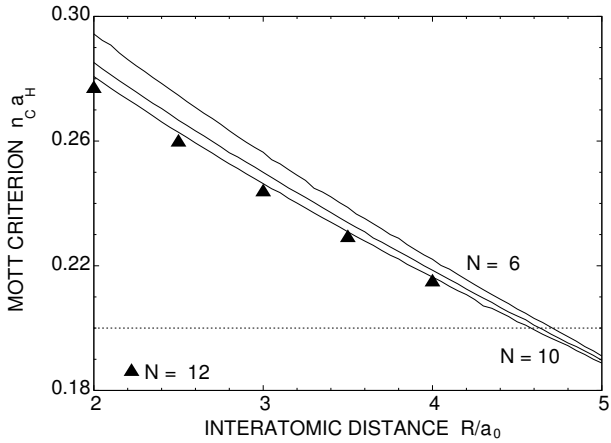} \par
\vspace{0.5cm}
\caption{
The Mott-criterion value $n_Ca_H$ vs.\ $R$ and for different
number of sites $N=6\div 10$. The horizontal line marks
the value for bulk systems.
\label{rys:ncah}}
\end{figure}

The determination of the basis $\{w_i({\bf r})\}_{i=1,\dots,N}$ allowed 
for determination of the interaction parameters 
(cf.\ Table~\ref{tab:ukvj}) and
the hopping integrals $t_m$ (cf.\ Table~\ref{tab:hopp}). 
Therefore, the bare bandwidth can be also defined as
\begin{equation}
\label{bandw}
  W=\max\{\epsilon_k\}-\min\{\epsilon_k\}.
\end{equation}
In Table~\ref{tab:eband} we present (for $N=8$) the effective orbital
size $a_H={\alpha}_{\min}^{-1}$ (in units of $a_0$), 
the bandwidth $W$, the effective interaction
parameters $U-K_1$, the $W/(U-K_1)$ ratio, and
the product of the carrier density $n_C=1/R$ and the optimal
orbital radius $a_H={\alpha}_{\min}^{-1}$. The product $n_Ca_H$ 
illustrates the Mott criterion for localization \cite{mott}, which can
be generalized first to  the case of $d$-dimensional infinite lattice,
which takes the form $n_C^{1/d}a_H\approx 0.2$. The critical
value of the product $0.2$ is reached for $R\approx 4.5 a_0$
and hence we should ask, whether this is a
coincidence or if it reflects the localization onset for
the correlated electrons in these nanoscopic systems.
This criterion is not strongly dependent on $N$, as shown 
in Fig.~\ref{rys:ncah} and hence represents an intrinsic property, 
independent of the system size for $N\geq 8$.

\begin{table}
\caption{Single-particle parameters versus interatomic distance
for $N=10$ sites.
\label{tab:hopp}}
\begin{tabular}{ r|r|r|r|r|r|r }
$R/a_0$ & $\epsilon_a$ & 
$t_1$ & $t_2$ & $t_3$ & $t_4$ & $t_5$ \\ 
 & [Ry] & [Ry] & [mRy] & [mRy] & [mRy] & [mRy] \\ \tableline 
2.0 & -4.461 & 
-0.585 & 87.0 & -8.93 & 1.29 & -0.413 \\ 
2.5 & -4.077 & 
-0.330 & 44.0 & -4.10 & 0.54 & -0.154 \\ 
3.0 & -3.712 & 
-0.200 & 23.6 & -2.00 & 0.23 & -0.006 \\ 
3.5 & -3.399 & 
-0.127 & 13.0 & -0.99 & 0.10 & -0.002 \\ 
4.0 & -3.138 & 
-0.083 & 7.5 & -0.54 & 0.05 & -0.009 \\ 
4.5 & -2.920 & 
-0.055 & 8.2 & -0.66 & 0.05 & -0.008 \\ 
5.0 & -2.737 & 
-0.037 & 4.3 & -0.28 & 0.02 & -0.002 \\ 
\end{tabular}
\end{table}

For that purpose we have calculated basic 
quantities signalling such a crossover, each of which 
can be characterized briefly under the following headings:
\begin{enumerate}
\setcounter{romenum}{0}

\refstepcounter{romenum}
\item[(\theromenum)]
\label{th:spin}
The total {\em spin length} per site. 
We characterize it by its square, i.e.\ by 
$\left<{\bf S}_i^2\right>=
\left<0\right|{\bf S}_i^2\left|0\right>$, where
${\bf S}_i=\left(S_i^+,S_i^-,S_i^z\right)=
\left(a_{i\uparrow}^{\dagger}a_{i\downarrow},
a_{i\downarrow}^{\dagger}a_{i\uparrow},
(n_{i\uparrow}-n_{i\downarrow})/2\right)$ 
is the electron spin operator for the atomic site $i$. 
It is easy to prove \cite{spawo} that 
\begin{equation}
\label{spinnn}
  \left<{\bf S}_i^2\right>=\frac{3}{4}
  \left(1-2\left<n_{i\uparrow}n_{i\downarrow}\right>\right).
\end{equation}
Therefore, in the atomic limit with one electron per atom we have that
$d^2\equiv\left<n_{i\downarrow}n_{i\uparrow}\right>=0$
and hence the spin length acquires the Pauli-spin limit 
$S(S+1)=3/4$. In the opposite, Hartree-Fock limit ($U_{\rm eff}\ll W$),
$\left<n_{i\uparrow}n_{i\downarrow}\right>=1/4$, and 
$\left<{\bf S}_i^2\right>=3/8$.
Hence, the quantity $\Theta_{\rm M}=(4/3)\left<{\bf S}_i^2\right>$
starts from the value $1/2$ for $R\rightarrow 0$ and approaches
the value of unity when the atomic-like localized states are more 
proper. Note that the ground state of the system is always a total spin 
singlet, i.e.\ $\left<0\right|\sum_{i=1}^N{\bf S}_i\left|\right>=0$.

\refstepcounter{romenum}
\item[(\theromenum)]
\label{th:nksig}
The dispersion of the statistical distribution of function
$n_{k\sigma}=\bra{0}a_{i\sigma}^{\dagger}a_{i\sigma}\ket{0}\equiv
\bra{0}\hat{n}_{k\sigma}\ket{0}$.
The distributions in both the optimized correlated state (i.e.\ with
$\alpha=\alpha_{\min}$) and in the state without such an optimization
(i.e.\ for $\alpha=a_0^{-1}$) is displayed in Fig.~\ref{rys:nks}ab. 
As the case with the optimized size $\alpha^{-1}=\alpha_{\min}^{-1}$
has a lower energy (see below), the behavior of $n_{k\sigma}$
confirms the existence of the Fermi ridge at $k=k_F=\pi/2R$
for $R\sim2a_0$ followed by its gradual diminution to its disappearance 
with the increasing $R$. 
The presence of the Fermi ridge speaks directly
in favour of delocalized (metallic) state of electrons \cite{migd}.
The distribution for $R\simleq 4a_0$ is essentially the Fermi-Dirac
distribution modified by the Fermi liquid effects, which is smeared out
completely for $R>5a_0$. 
To put this process on the 
quantitive grounds we have calculated first the dispersion
of the statistical distribution  
$$
  \sigma^2\{n_{k\sigma}\}=
  \frac{1}{2N}\sum_{k\sigma}
  \left<0\right|\hat{n}_{k\sigma}\left|0\right>^2
$$
\begin{equation}
\label{nksig}
  -\left(\frac{1}{2N}\sum_{k\sigma}
  \left<0\right|\hat{n}_{k\sigma}\left|0\right>\right)^2.
\end{equation}
In the Hartree-Fock limit this quantity can be calculated 
by assuming that 
$\left<\hat{n}_{k\sigma}\right>=\Theta\left(\mu-\epsilon_k\right)$, 
so that $\sigma^2\{n_{k\sigma}\}=1/4$, 
whereas it vanishes for the even momentum distribution 
$(n_{k\sigma}=1/2)$, when the particle position 
is sharply defined on atom, i.e.\ for the localized states reducing
to the atomic states. 

\refstepcounter{romenum}
\item[(\theromenum)]
\label{th:spinspin}
The nearest neighbor {\em spin-spin correlation function}
$\Theta_{\rm AF}=-\left<{\bf S}_i\cdot{\bf S}_{i+1}\right>$.
It should be zero in the Hartree-Fock limit and reach the
value $(3/4)$ for the nearest-neighbor singlet state which mimics
the antiparallel orientation of the classical spins. 

\refstepcounter{romenum}
\item[(\theromenum)]
\label{th:nksz}
The {\em Fermi discontinuity {\em (ridge)} disappearance}. 
It is defined as \cite{migd}
\begin{equation}
\label{nksz}
  \Delta n_{k_F}=n_{k=k_F-0}-n_{k=k_F+0}.
\end{equation}
This difference has been interpolated by a parabola 
$n_{k\sigma}=\alpha k^2+\beta k+\gamma$ 
from both sides leading to the value $\Delta n_{k_F}^{\rm par}$ at the jump. 
Such an interpolation simulates a quasicontinuous function 
$n_{k\sigma}$, which would be present for large $N$. 
The resultant $R$ dependences of both computed and interpolated 
$\Delta n_{k_F}$ values for $N=10$ 
are listed in Table~\ref{tab:nks}. The discontinuity disappears in the 
range of $R_c=4\div 4.5 a_0$. The large uncertainty is due to the poor
statistics of the points (3 points on each side).  
However, it 
is close to the values deducted from $R$ dependence of the quantities 
(\ref{th:spin})-(\ref{th:spinspin}), as we discussed next.
\end{enumerate}

\begin{figure}
\centering\epsfxsize=8cm\epsffile{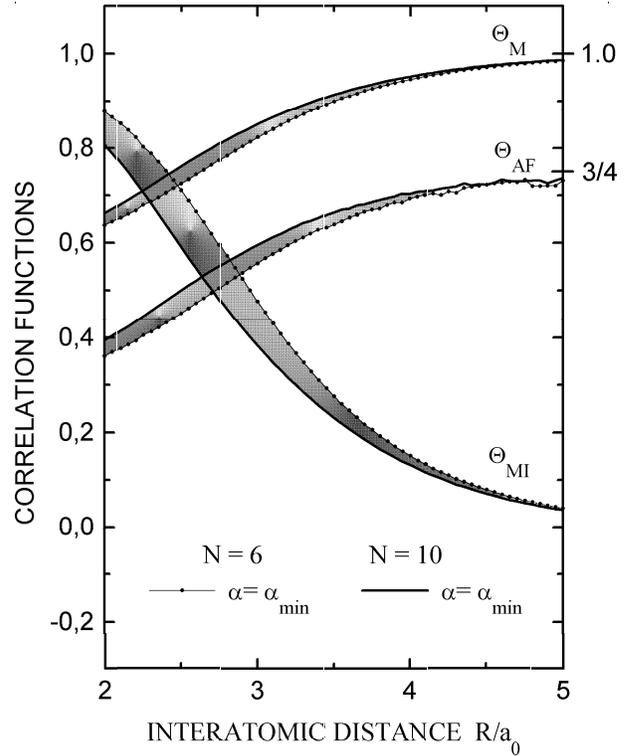} \par
\vspace{0.5cm}
\caption{
Correlation functions versus distance $R$, 
characterizing the crossover from itinerant
to localized state. The shaded areas are
drawn to illuminate the results convergence
with the increasing $R$.
\label{rys:th}}
\end{figure}

The quantities characterized (\ref{th:spin})-(\ref{th:spinspin}) are
displayed in Fig~\ref{rys:th}, 
where we have shadowed the difference between the results
for $N=6$ and $10$ (the results for $N=8$ fall in the area)
to amplify the convergence of the numerical results.
We see that the quantities $\Theta_{\rm MI}$, $\Theta_{\rm AF}$,  
and $\Theta_{\rm M}$ acquire the atomic 
values within the $5\%$ range for $R/a_0\approx 5$, 
which corresponds to the interatomic distance $R\approx 2.6\AA$. It
should be underlined that we do not expect any discontinuous phase 
transition for this finite (in fact, nanoscopic!) system. However, 
with the help of the characteristics provided above one can 
define an experimental criterion of localization (here we 
suggest that the $5\%$ margin for the characteristics 
to fall within the atomic limit values, is a natural one). Also, all the 
characteristics defined above are defined from the side of
the delocalized (metallic) state. To define properly the Mott insulating
state as that of a Heisenberg magnet (i.e.\ with frozen orbital
degrees of freedom), we have to consider the spin-spin correlation
function directly, as well as the magnitude of
the superexchange (kinetic exchange), both as a function of
interatomic distance. 

\begin{table}
\caption{
Effective Bohr radius ($a_H$), the bare bandwith ($W$), 
the interaction parameter ($U-K_1$), the bandwith to 
interaction ratio, and the Mott criterion ($n_Ca_H$), respectively.
\label{tab:eband}}
\begin{tabular}{ r|r|r|r|r|r }
$R/a_0$ & $a_H/a_0$ & $W$ & $U-K_1$
 & $W/(U-K_1)$ & $n_Ca_H$ \\ \tableline  
2.0 & 0.570 & 2.381 & 1.224
 & 1.945 & 0.285 \\
2.5 & 0.667 & 1.340 & 1.106
 & 1.212 & 0.267 \\ 
3.0 & 0.750 & 0.810 & 1.026
 & 0.790 & 0.250 \\ 
3.5 & 0.818 & 0.512 & 0.976
 & 0.525 & 0.234 \\ 
4.0 & 0.874 & 0.333 & 0.944
 & 0.353 & 0.219 \\ 
5.0 & 0.948 & 0.148 & 0.925
 & 0.160 & 0.190 \\ 
\end{tabular}
\end{table}

The Fermi ridge is sharply defined in the many-fermion system only when 
the perturbation expansion is convergent (cf.\ Luttinger \cite{migd}). 
It is certainly not at the Mott metal-insulator transition. Our results show
that the Fermi discontinuity disappears with the increasing lattice 
parameter even in the situation when we have a crossover transition
from a metal to an insulator. 
This means that low-dimensional (finite) systems
cannot always be analyzed perturbationally even thought they do not
exhibit phase transformation in the thermodynamic sense. 
Obviously, part of the jump at the Fermi points is due to the discretness
of the particle quasimomentum in this finite-size system.

\section{Kinetic exchange and spin-spin correlations}
\label{kin}

\begin{figure}
\centering\epsfxsize=8cm\epsffile{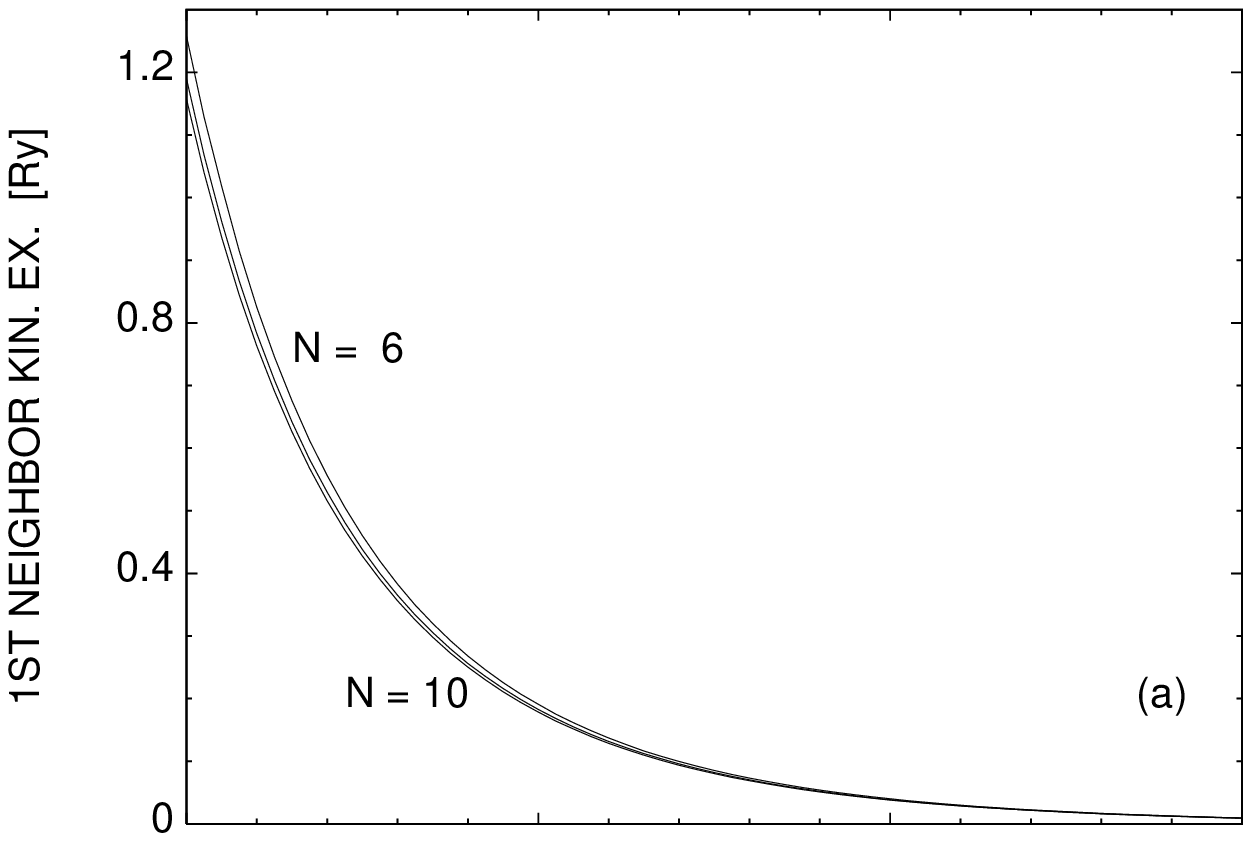} \propar
\centering\epsfxsize=8cm\epsffile{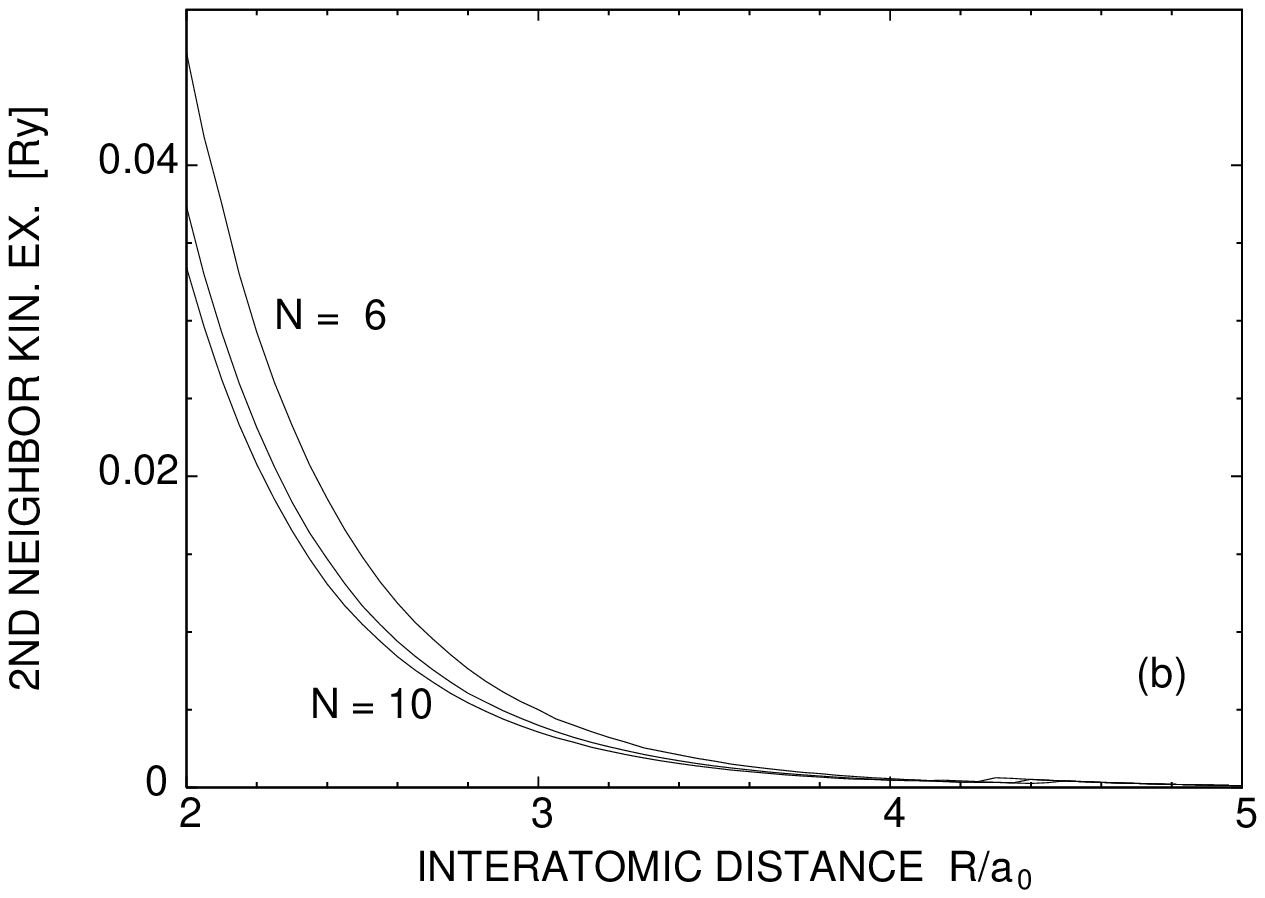} \par
\vspace{0.5cm}
\caption{
$R$ dependence of the kinetic exchange integrals $J^{(1)}_{\rm kex}$ 
(a) and $J^{(2)}_{kex}$ (b) for different $N=6\div 10$.
The results for the dominant integral (a) are weakly $N$ dependent. 
\label{rys:jkex}}
\end{figure}

In our system the total spin is conserved. Since the 
ground state is a spin singlet, we have that 
$\bra{0}\left(\sum_{i=1}^{N}{\bf S}_i\right)^2\ket{0}=0$. 
From this we can derive the sum rule for the ground state of the
Heisenberg system in the form
\begin{equation}
\label{spinsum}
  \left<{\bf S}_i^2\right>+\sum_{m=1}^{N-1}
  \left<{\bf S}_i\cdot{\bf S}_{i+m}\right>=
  \frac{1}{N}S_{\rm tot}\left(S_{\rm tot}+1\right)=0. 
\end{equation}
In our case the spins are defined in the Fock space (see above).
Therefore, the sum rule may not be obeyed, since the spin magnitude 
is smaller. This is explicitly evident in Table~\ref{tab:s0sn},
where the spin-spin correlation functions 
$\left<{\bf S}_i\cdot{\bf S}_{i+m}\right>$ are listed as a function 
of $R$. The long-range correlations set in and oscillate in sign as
the atomic limit is approached . 
This feature of the correlation function must be induced 
by the Anderson antiferromagnetic kinetic exchange interaction. 
We have calculated the kinetic exchange 
integrals between $m$-th neighbors defined as \cite{ande}
\begin{equation}
\label{jkex}
  J_{\rm kex}^{(m)}=\frac{4\left(t_m+V_m\right)^2}{U-K_m}, 
\end{equation}
and displayed them in Table~\ref{tab:jkex}. The total strength
of kinetic exchange ($\approx\sum_{m=1}^{3}J_{\rm kex}^{(m)}$)
has also been compared with the corresponding quantity 
($\approx\sum_{m=1}^{3}J_m$) expressing the strength
of the Heisenberg-Dirac exchange (the former dominates in 
the full $R$ range listed). Obviously, Eq.~(\ref{jkex}) provides,
strictly speaking, the reliable values for the kinetic 
exchange integral only in the limit $U-K_m\gg\left|t_m+V_m\right|$. 
All the integrals are of antiferromagnetic character, since
they express virtual hopping processes
between $m$-neighbors, which occur (and diminish the 
system energy in the second order) only when spins on 
the two sites are oriented antiparallel. The results for 
the integrals $J_{\rm kex}^{(m)}$ are only weakly dependent on $N$, as
exemplified in Fig.~\ref{rys:jkex}ab, where the first two exchange 
integrals have been shown for $N=6,8,10$. 
The kinetic exchange is rather strong as
for e.g.\ $R=2.65\AA$ we have that 
$J_{\rm kex}^{(1)}=0.124\mbox{ eV}$ ($\approx 1440\mbox{ K}$), 
$J_{\rm kex}^{(2)}=1.6\mbox{ meV}$ ($\approx 19\mbox{ K}$), 
$J_{\rm kex}^{(3)}=0.44\mbox{ meV}$ ($\approx 5\mbox{ K}$).
Such a strong superexchange is observed only 
in the two-dimensional superconducting cuprates \cite{birg}.

\begin{figure}
\centering\epsfxsize=8cm\epsffile{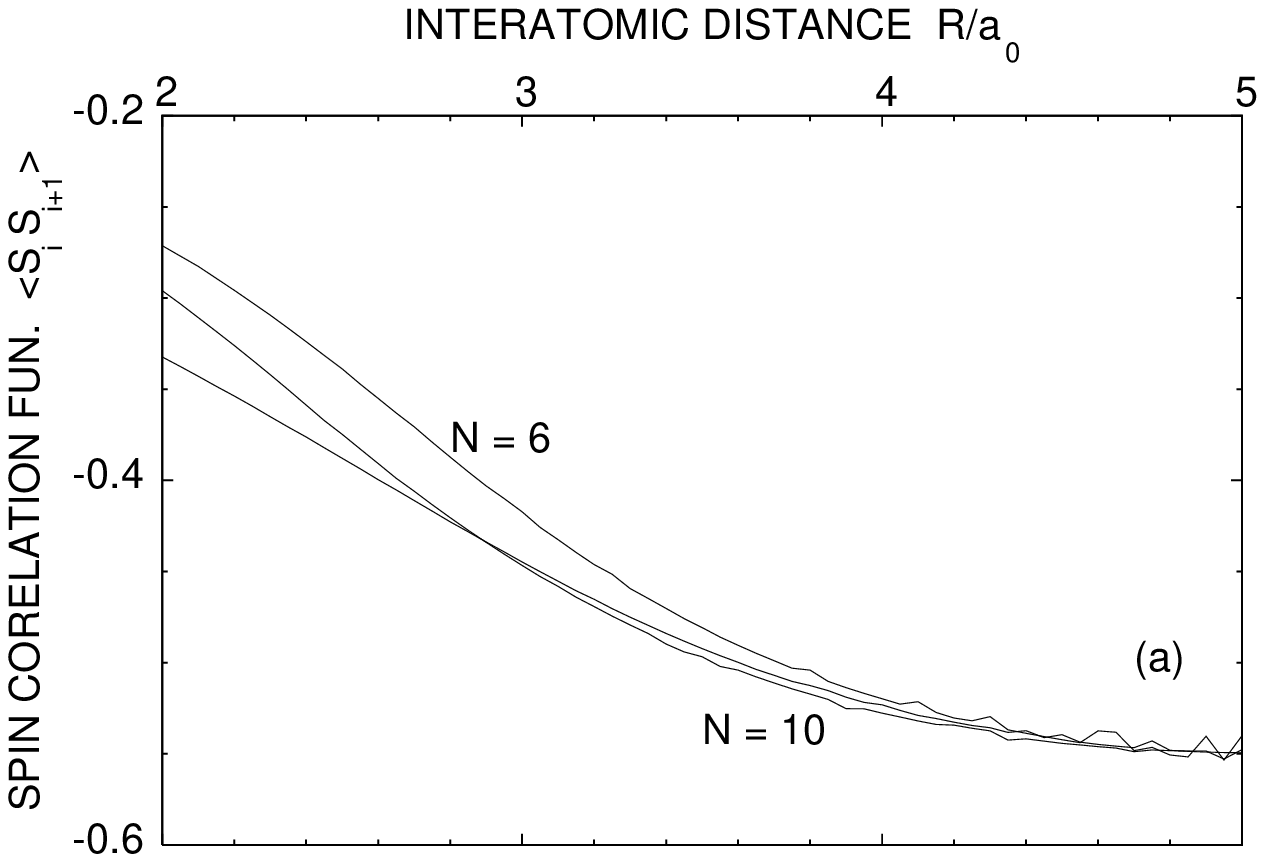} \propar
\centering\epsfxsize=8cm\epsffile{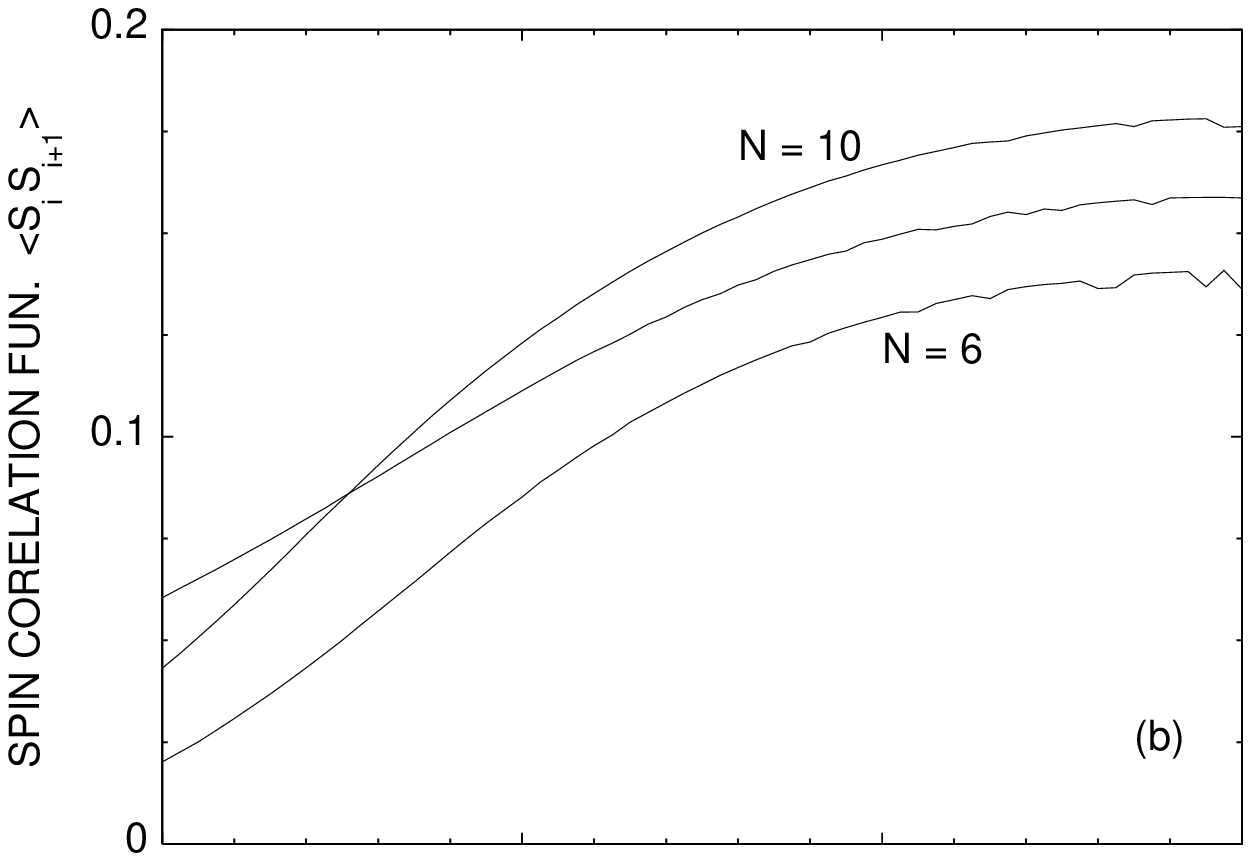} \propar
\centering\epsfxsize=8cm\epsffile{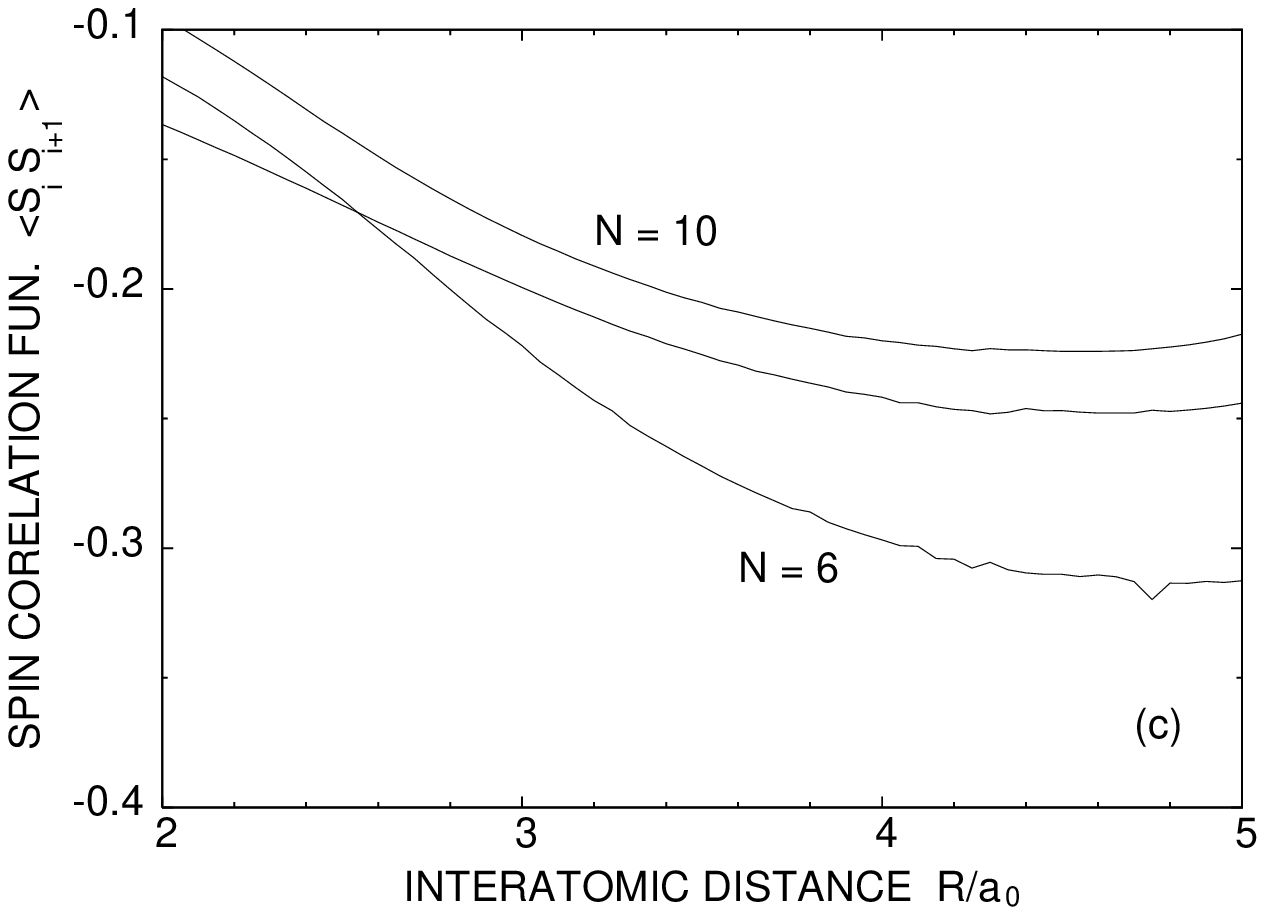} \par
\vspace{0.5cm}
\caption{
Distance dependence of the spin-spin correlation functions
$\left<{\bf S}_i\cdot{\bf S}_{i+p}\right>$ 
for $p=1\div 3$ and $N=6\div 10$. Note the oscillatory 
character with $p$ expressing the antiferromagnetic correlations. 
\label{rys:s0sn}}
\end{figure}

The strong system-size dependence of the spin-spin correlation 
functions is shown in Fig.~\ref{rys:s0sn}a-c. This is not strange, 
since the spins correlate over much longer distance than the range of 
the interactions as express collective properties of the system 
(the same can be said about the Fermi discontinuity $\Delta n_{k_F}$).

\begin{table}
\caption{
The $R$ dependence of the Fermi discontinuity
$\Delta n_{k_F=\pi/2R}$ and the value obtained by the 
parabolic interpolation $\Delta n^{par}_{k_F}$.
\label{tab:nks}}
\begin{tabular}{ r|r|r }
$R/a_0$ 
 & $\Delta n_{k_F}$ & $\Delta n_{k_F}^{\rm par}$ \\ \tableline
2.0 & 0.8264 & 0.7326 \\ 
2.5 & 0.6148 & 0.4244 \\ 
3.0 & 0.4033 & 0.1582 \\ 
3.5 & 0.2645 & 0.0358 \\ 
4.0 & 0.1791 & -0.0040 \\ 
4.5 & 0.1232 & -0.0140 \\ 
5.0 & 0.0857 & -0.0130 \\ 
\end{tabular}
\end{table}

\begin{table}
\caption{
The spin-spin correlation functions 
$\left<{\bf S}_i\cdot{\bf S}_{i+p}\right>$
for $p=1,2,3$ and $N=10$, as a function of lattice constant.
\label{tab:s0sn}}
\begin{tabular}{ r|r|r|r }
$R/a_0$
 & $\left<{\bf S}_i\cdot{\bf S}_{i+1}\right>$ 
 & $\left<{\bf S}_i\cdot{\bf S}_{i+2}\right>$ 
 & $\left<{\bf S}_i\cdot{\bf S}_{i+3}\right>$  \\ 
\tableline
2.0 & -0.2959 & 0.0431 & -0.0951  \\ 
2.5 & -0.3748 & 0.0845 & -0.1399  \\ 
3.0 & -0.4467 & 0.1230 & -0.1793  \\ 
3.5 & -0.4968 & 0.1501 & -0.2052  \\ 
4.0 & -0.5277 & 0.1668 & -0.2200  \\ 
4.5 & -0.5442 & 0.1754 & -0.2241  \\ 
5.0 & -0.5477 & 0.1762 & -0.2175  \\ 
\end{tabular}
\end{table}

\begin{table}
\caption{
Distance dependence of kinetic exchange integrals
$J^{(m)}_{\rm kex}$ for $m=1,2,3$, as well as the total kinetic
exchange and the summary direct exchange integral ($\sum J_m$).
\label{tab:jkex}}
\begin{tabular}{ r|r|r|r|r|r }
$R/a_0$ 
 & $J_{kex}^{(1)}$ & $J_{kex}^{(2)}$ & $J_{kex}^{(3)}$
 & $\sum J_{kex}^{(m)}$ & $\sum J_m$ \\
 & [Ry] & [mRy] & [mRy] & [Ry] & [mRy]  \\ \tableline   
2.0 & 1.190 & 37.34 & 0.493 & 1.227 & 11.98 \\
2.5 & 0.438 & 11.67 & 0.129 & 0.450 & 8.37 \\
3.0 & 0.183 & 3.99 & 0.036 & 0.187 & 5.99 \\
3.5 & 0.082 & 1.39 & 0.010 & 0.084 & 4.25 \\
4.0 & 0.039 & 0.48 & $2.7\cdot 10^{-3}$
 & 0.039 & 2.96 \\
4.5 & 0.019 & 0.42 & $2.5\cdot 10^{-3}$
 & 0.019 & 1.99 \\
5.0 & $9.3\cdot 10^{-3}$ & 0.12 & $4.4\cdot 10^{-4}$
 & $9.4\cdot 10^{-3}$ & 1.27 \\ 
\end{tabular}
\end{table}

\section{Electron-lattice interaction, zero-point motion 
	and dimerization from the first principles}
\label{elec}

\subsection{General form of local electron-phonon coupling}

\begin{figure}
\centering\epsfxsize=8cm\epsffile{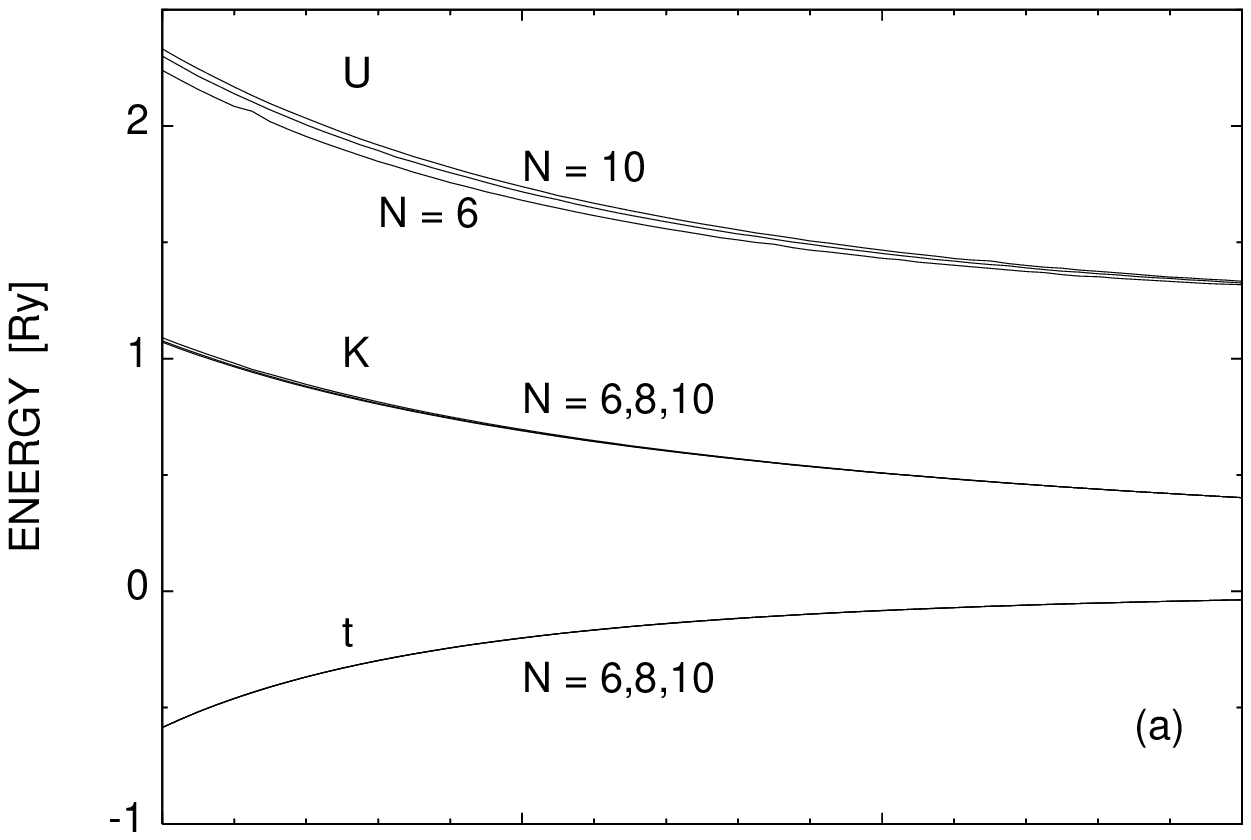} \propar
\centering\epsfxsize=8cm\epsffile{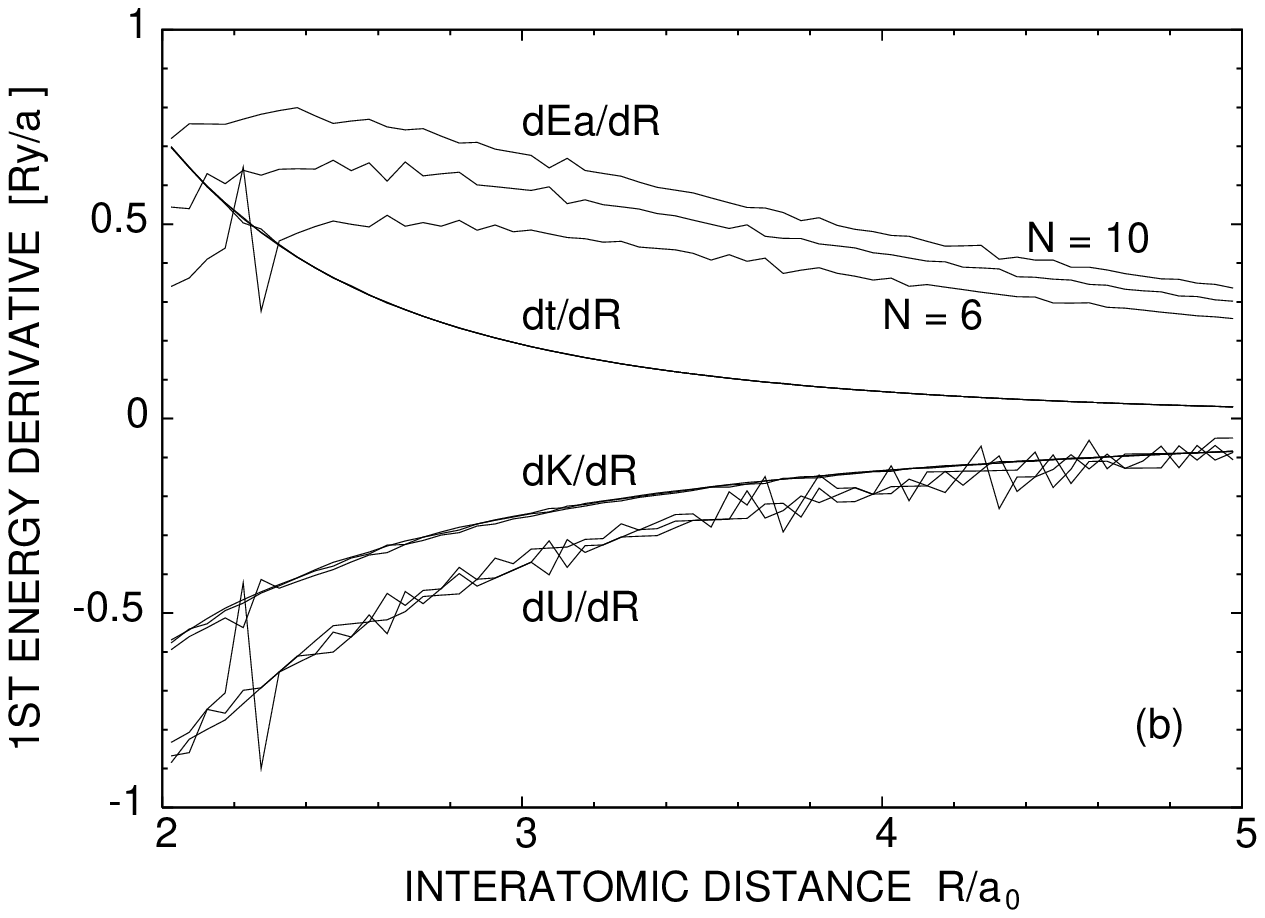} \par
\vspace{0.5cm}
\caption{
The interaction parameters (a) and their derivatives (b)
with respect to the lattice-parameter change; 
the latter  express the coupling-constant 
strength for different local electron-lattice couplings.
\label{rys:elph}}
\end{figure}

We can extend our method to include the local electron-lattice
coupling \cite{ovch}. In the second-quantized Hamiltonian of general
form (\ref{ham:lanext}) the positions of nuclei 
$\left\{{\bf R}_i\right\}_{i=1,\dots,N}$ 
are regarded as fixed, 
i.e.\ the ions are regarded as classical objects. 
If their positions are subject to a local shifts
$\left\{\delta{\bf R}_i\right\}_{i=1,\dots,N}$, 
then the Hamiltonian will change by the amount $\delta{\cal H}$,
so that now ${\cal H}\rightarrow {\cal H}+\delta{\cal H}$. 
The amount of the change can be calculated as
\begin{equation}
\label{delh}
  \delta{\cal H}\equiv
  \sum_{i=1}^N\frac{\delta{\cal H}}{\delta{\bf R}_i}
  \cdot\delta{\bf R}_i\equiv
  \sum_{i=1}^N\nabla_i{\cal H}\cdot\delta{\bf R}_i,
\end{equation}
which holds true for $|\delta {\bf R}|\ll R$.
Primarily, the nuclei shifts $\left\{\delta{\bf R}_i\right\}$ is felt by
the potential energy $V\left({\bf r}\right)\equiv
\sum_iV\left({\bf r}-{\bf R}_i\right)$
and by the Wannier functions $\left\{w_i\left({\bf r}\right)=
w_i\left({\bf r}-{\bf R}_i\right)\right\}$. 
These changes, in turn, induce the alteration of the microscopic
parameters ${\epsilon}_a$, $U$, $K_{ij}$, etc. In result, we can write
the Hamiltonian change in the form
$$
  \delta{\cal H}=
  \sum_i\left.\frac{\delta{\epsilon}_a}{\delta{\bf R}_i}\right|_0
  \cdot\delta{\bf R}_i\, n_i +
$$
$$
  {\sum_{ij\sigma}}'
  \left.\frac{\delta t_{ij}}{\delta{\bf R}_i}\right|_0
  \left(\delta{\bf R}_i-\delta{\bf R}_j\right)
  a_{i\sigma}^{\dagger}a_{j\sigma}
  +\sum_i\left.\frac{\delta U}{\delta{\bf R}_i}\right|_0
  \cdot\delta{\bf R}_in_{i\uparrow}n_{i\downarrow}
$$
\begin{equation}
\label{delhexp}
  +{\sum_{ij}}'\left.\frac{\delta K_{ij}}{\delta{\bf R}_i}\right|_0
  \cdot\left(\delta{\bf R}_i-\delta{\bf R}_j\right)n_in_j+\dots.
\end{equation}
One should note that we have not included explicitly the more distant
displacements, i.e.\ the terms of the type
$$
  {\sum_{ij}}'\left.\frac{\delta\epsilon_a}{\delta{\bf R}_j}\right|_0
  \cdot\delta{\bf R}_jn_i,\ \ \
  {\sum_{\stackrel{ij\sigma}{k\neq (i,j)}}}'
  \left.\frac{\delta t}{\delta{\bf R}_k}\right|_0
  \cdot\delta{\bf R}_ka_{i\sigma}^{\dagger}a_{j\sigma},\ \ \
  \mbox{etc.} 
$$
The last terms represent a coupling to more distant
ions on the atomic level position, or on the nearest-neighbor hopping, etc.
They are not included as we would like to determine first the
derivatives $\delta/\delta{{\bf R}_i}(\dots)$ 
from our above first-principle results 
(the subscript ``$0$'' means they are calculated for the periodic
arrangement of ions). 
In Fig.~\ref{rys:elph} (bottom panel) we display the corresponding 
derivatives, which play the role of the local electron-lattice coupling 
constants, as a function of the interatomic distance. 
The scattering of the points is caused by the numerical differentiation. 
For the sake of completeness, we have plotted in the top panel the
microscopic parameters vs.\ $R$, for $N=6\div 10$ atoms. 

There are few unique features of these results, which we would 
like to elaborate on. First, the interaction terms $\sim\delta U$ 
and $\sim\delta K$ diminish the system energy when the system distorts. 
Also, $dU/dR\sim -d\epsilon_a/dR$, 
but effectively $\left|dK/dR\right|$ overcomes $dt/dR$, 
so that the net effects in the 
insulating state $\left\{\hat{n}_i=1\right\}$ favours the system 
distorsion (note a rather weak dependence on $N$). Furthermore, 
the coupling constants are relatively large. For example, 
$\lambda_a/\left|\delta R\right|
\equiv d\epsilon_a/dR\approx 0.5\mbox{ Ry}/a_0$
we obtain for the distorsion $\left|\delta R\right|/R\approx 0.1$
(estimating the of zero-point motion amplitude; see below)
the value of $\lambda_a\approx 2\mbox{ eV}$
for $R/a_0=3$ (i.e.\ $R=1.6\AA$). Similarly 
$\lambda_U\equiv\left(dU/dR\right)\left|\delta R\right|
\approx -2.4\mbox{ eV}$, 
$\lambda_K\approx -1.6\mbox{ eV}$, $\lambda_t\approx 0.7\mbox{ eV}$. 
They represent a sizeble fraction of the value 
$U-K_1\approx 13.8\mbox{ eV}$. 
What is more important, $\lambda_U/W\approx 0.25$.
However, the kinetic exchange integral is 
$J_{\rm kex}^{(1)}\approx 2.5\mbox{ eV}$, 
a value close to $\left|\lambda_U\right|$.
This means that for the linear chain of hydrogen atoms 
the electron-lattice and magnetic interactions are of comparable
magnitude. In the system with heavier ions the electron-lattice
coupling should be divided roughly by $M_i^{1/2}$, where $M_i$ is
the ion mass. Also, in the insulating state the $U$ and $K$ 
parameters (and their derivatives) should be roughly diminished
by the relative dielectric constant $\epsilon$ of the medium. 
Roughly, $\left(M_i/M_H\right)^{1/2}\sim 5\div 10$, and also 
$\epsilon\sim 5\div 10$, so that
the proportions between the above parameters should remain
of the same order even though their absolute values diminish by the 
factor $5\div 10$. But this means that the analysis of the 
strongly correlated low-dimensional systems should include on equal 
footing both local electron-electron and electron-lattice couplings. 
There are quite few relevant parameters 
$\left(\epsilon_a,W,U,K_m,\lambda_a,\lambda_K,\lambda_U\right)$
in that situation,
so the usual approach of solving Hamiltonian ${\cal H}+\delta {\cal H}$ 
by regarding all those quantities 
as free parameters of the model, does not look promising. 
Our first-principle approach determines the value of the parameters
accurately so their values are known for fixed R. 
However, needless to say that our method must be extended to a more 
realistic situation involving e.g.\  $\left({\rm CuO}_2\right)_n$ 
planar clusters to be applicatable to the high-$T_c$ systems. 

The solution of the Hamiltonian incorporating the electron-lattice 
interactions requires a separate analysis and will not be discussed here. 
In the remaining part, we concentrate only on the evolution of the  zero-point 
motion and of the dimerization as a function of interatomic distance.

\subsection{Zero-point motion}

In the harmonic approximation the acoustic phonons for the linear chain
have the dispersion relation of the form
\begin{equation}
\label{omph}
  \omega_k=
  2\left(\frac{C}{M}\right)^{\half}\sin\left(\frac{\pi k}{N}\right),
\end{equation}
where $M$ is the ion mass (we take the proton mass here) and $C$ is
the elastic constant, which can be calculated for longitudinal modes
from the differentiation of the total ground state energy $E_G$ 
(with inclusion of the interionic interactions), namely
\begin{equation}
  \label{cspr}
  C=\frac{1}{N}\frac{\partial^2E_G}{\partial^2R}.
\end{equation}
One should note that due to the global instability 
$(\partial E_G/\partial R<0)$ 
we should place the system in a box stabilizing it (e.g.\ the system 
represents a linear ring on a substrate stabilizing its geometry).
Also, the $k=0$ is a Goldstone mode, so we select the values 
of $k=1,2,\dots,N-1$, i.e.\ chose the center-of-mass reference
system. In result, the contribution of zero-point motion to the
system energy is
\begin{equation}
  \label{degph}
  \Delta E_G^{\rm ph}=\sum_k\half\hbar\omega_k=
  \frac{\hbar}{M^{\half}}
  \left(\frac{1}{N}\frac{\partial^2 E_G}{\partial R^2}\right)^{\half}
  \sum_{k=1}^{N-1}\sin\left(\frac{\pi k}{N}\right).
\end{equation}
In the atomic units, it takes the form
\begin{equation}
  \label{degphau}
  \Delta E_G^{\rm ph}\eqau\left(\frac{2m}{M}\right)^{\half}
  \left(\frac{1}{N}\frac{\partial^2 E_G}{\partial R^2}\right)^{\half}
  \sum_{k=1}^{N-1}\sin\left(\frac{\pi k}{N}\right).
\end{equation}

To estimate the amplitude $\Delta R$ of zero-point motion we note 
that quasiclassically we can write for the individual normal
mode that
\begin{equation}
  \label{drphintro}
  \frac{1}{2}M\omega_k^2\left(\Delta R_k\right)^2=
  \frac{1}{2}\hbar\omega_k,
\end{equation}
where $\Delta R_k$ is the classical amplitude of the vibrations
associated with $k$-th mode. Introducing the global classical
amplitude
\begin{equation}
  \label{drph}
  \left(\Delta R\right)^2=\frac{1}{N}\sum_{k=1}^{N-1}
  \frac{\hbar}{M\omega_k}
\end{equation}
we obtain in atomic units that
\begin{equation}
  \label{drphau-lanext}
  \left(\Delta R\right)^2\eqau
  \frac{1}{N}\left(\frac{m}{2M}\right)^{\half}
  \left(\frac{1}{N}\frac{\partial^2 E_G}{\partial R^2}\right)^{-\half}
  \sum_{k=1}^{N-1}\frac{1}{\sin(\pi k/N)}.
\end{equation}
Approximating the summation by integration we have that 
$\left(\Delta R\right)^2$ 
is divergent and $\left(\Delta R\right)^2\sim\ln N$. The above formula
expresses the first-order contribution to the lattice dynamics. 
This contribution \cite{anis} appears on the top of the 
optimized energy $E_G\equiv E_G\left(\alpha=\alpha_{\min},R\right)$.

\subsection{Lattice dimerization: Basis halving}

\begin{figure}
\centering\epsfxsize=8cm\epsffile{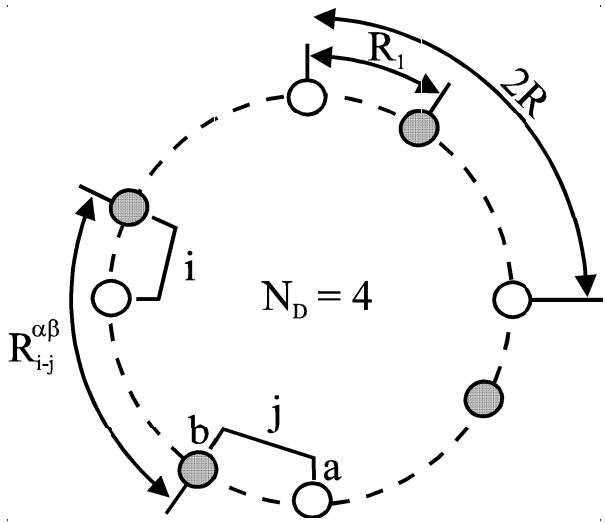} \par
\vspace{0.5cm}
\caption{
Schematic representation of the chain dimerization with
the characteristic distance and the distorsion into
two sublattices.
\label{rys:dimm}}
\end{figure}

It is well known that one-dimensional systems are
instable with respect to the dimerization (the Peierls distorted
phase \cite{peie}). We incorporate the 
distorted phase into our rigorous analysis of the ground 
state properties. For that purpose, we define two sets
of atomic basis functions $A$ and $B$ corresponding to even and odd
lattice sites, shown schematically in Fig.~\ref{rys:dimm}, namely
$$
  \Psi^A_i({\bf r})=\Psi_{2i}({\bf r}),\ \ \ \  
  \mbox{and}\ \ \ \ 
  \Psi^B_i({\bf r})=\Psi_{2i+1}({\bf r}).
$$
The index for each sublattice takes $N_D\equiv N/2$ values 
(this is the {\em basis halving}). 
We construct next the two sets of the Bloch functions starting from this
atomic basis, which are
\begin{equation}
  \label{blocha:dimm}
  \Phi^A_k=\frac{{\cal N}^A_k}{N_D^{1/2}}
  \sum_{j=0}^{N_D-1}\Psi^A_j\exp\left(i\frac{2\pi kj}{N_D}\right),
\end{equation}
\begin{equation}
  \label{blochb:dimm}
  \Phi^B_k=\frac{{\cal N}^B_k}{N_D^{1/2}}
  \sum_{j=0}^{N_D-1}\Psi^B_j\exp\left(i\frac{2\pi kj}{N_D}\right),
\end{equation}
where the quantum number $k=0,\dots,N_D-1$ enumerates the
points in the reduced zone ($N_D=N/2$), and the normalization 
factors are
\begin{equation}
  \label{norm:dimm}
  {\cal N}_k={\cal N}^A_k={\cal N}^B_k=\left[
  \sum_{p=0}^{N_D-1}S^{AA}_p\cos\left(\frac{2\pi kp}{N_D}\right)
  \right]^{-1/2}.
\end{equation}
The overlap integrals $S_p$ are calculated from the prescription:
$S_{i-j}^{\alpha\beta}=S\left(R_{ij}^{\alpha\beta}\right)$,
where $\alpha,\beta=A$ or $B$, and $i-j=0,\dots,N_D-1$. 
Obviously, $R_{ij}^{\alpha\beta}=R_{ji}^{\beta\alpha}$ 
and $R_{ij}^{AA}=R_{ij}^{BB}$; the same symmetry is obeyed by 
all quantities dependent on $R_{ij}^{\alpha\beta}$.

Each sublattice contains only half of the total 
number of lattice sites. This circumstances leads to the
nonorthogonality of the Bloch functions 
$\left<\Phi_k^A\Phi_k^B\right>=S_{AB}(k)\neq 0$, 
even though we had before
$\left<\Phi_k\Phi_{k'}\right>=\delta_{kk'}$. 
Therefore, the orthogonalized wave functions are
\begin{equation}
  \label{orto:dimm}
  \Phi^1_k=\beta\left(\Phi^A_k+\gamma^*_k\Phi^B_k\right),
  \ \ \ \ 
  \Phi^2_k=\beta\left(\Phi^B_k+\gamma_k\Phi^A_k\right),
\end{equation}
with
\begin{equation}
  \label{gam:dimm}
  \gamma_k=-\frac{S_{AB}(k)}{1+\sqrt{1-|S_{AB}(k)|^2}},
\end{equation}
and
$$
  \beta_k=\left[1+\frac{|S_{AB}(k)|^2}{\left(1+
    \sqrt{1-|S_{AB}(k)|^2}\right)^2}\right.
$$
\begin{equation}
\label{bet:dimm}
  -\left.
  \frac{|S_{AB}(k)|^2}{1+\sqrt{1-|S_{AB}(k)|^2}}\right]^{-1/2}.
\end{equation}
The overlap integral $S_{AB}(k)$ for the Bloch functions can be
related to the overlaps $S_p^{AB}$ between $p$-th neighbors
located on different sublattices, namely
$$
  S_{AB}(k)={\cal N}_k\sum_p\left[\frac{S^{AB}_p+S^{BA}_p}{2}
  \cos\left(\frac{2\pi kp}{N_D}\right)\right.
$$
\begin{equation}
\label{sab:dimm}
  +\left.
  i\frac{S^{AB}_p-S^{BA}_p}{2}\sin\left(\frac{2\pi kp}{N_D}\right)
  \right].
\end{equation}
In effect, the expansion of the Wannier function in the 
basis of atomic functions has the form
\begin{equation}
  \label{wann:dimm}
  w^{\alpha}_i=\sum_{i\beta}\beta^{\alpha\beta}_{i-j}\Psi^{\beta}_j,
\end{equation}
Taking the inverse (space) Fourier transforms of the orthogonalized
Bloch functions we obtain the coefficients 
$\beta_{i-j}^{\alpha\beta}\equiv\beta_p^{\alpha\beta}$
in the form
\begin{equation}
  \label{betaa:dimm}
  \beta^{AA}_p=N_D^{-1}\sum_k{\cal N}_k\beta_k
  \cos\left(\frac{2\pi kp}{N_D}\right)=\beta^{BB}_p,  
\end{equation}
$$
  \beta^{AB}_p=N_D^{-1}\sum_k{\cal N}_k\beta_k\left[
  \mbox{Re}\gamma_k\cos\left(\frac{2\pi kp}{N_D}\right)\right.
$$
\begin{equation}
  \label{betab:dimm}
  +\left.
  \mbox{Im}\gamma_k\sin\left(\frac{2\pi kp}{N_D}\right)\right],
\end{equation}
$$
  \beta^{BA}_p=N_D^{-1}\sum_k{\cal N}_k\beta_k\left[
  \mbox{Re}\gamma_k\cos\left(\frac{2\pi kp}{N_D}\right)\right.
$$
\begin{equation}
  \label{betba:dimm}
  -\left.
  \mbox{Im}\gamma_k\sin\left(\frac{2\pi kp}{N_D}\right)\right].
\end{equation}
Note that $\beta_p^{AB}=\beta_{-p}^{BA}$, 
in accordance with the definition 
of the relative distance $R_{ij}^{\alpha\beta}$ defined in 
Fig.~\ref{rys:dimm}. 

With the help of the orthonormal basis 
$$
  \left\{w_i^{\alpha}\right\}_{\stackrel{\scriptstyle\alpha=A,B}%
{i=0,\dots,N_D-1}},
$$ 
we can define the system Hamiltonian with
inclusion of {\em all} two-site interactions and {\em all} hopping
properties in the following manner
$$
  {\cal H}=\sum_{i=0}^{N_D-1}\sum_{\alpha=A,B}
  \left\{\epsilon_an_{i\alpha}+
  Un_{i\alpha\uparrow}n_{i\alpha\downarrow}\right.
$$
$$
  +\sum_{j\beta<i\alpha}\left[\left(K_{i-j}^{\alpha\beta}
  -\frac{1}{2}J_{i-j}^{\alpha\beta}\right)n_{i\alpha}n_{j\beta}
  -2J_{i-j}^{\alpha\beta}{\bf S}_{i\alpha}{\bf S}_{j\beta}
  \right.
$$
$$
  +\sum_{\sigma}\left(t_{i-j}^{\alpha\beta}
  +V_{i-j}^{\alpha\beta}(n_{i\alpha\bar{\sigma}}+
  n_{j\beta\bar{\sigma}})\right)\left(
  a^{\dagger}_{i\alpha\sigma}a_{j\beta\sigma}+\mbox{h.c.}\right)+
$$
\begin{equation}
  \label{ham:dimm}
  \left.\left.
  +J_{i-j}^{\alpha\beta}\left(a^{\dagger}_{i\alpha\uparrow}
  a^{\dagger}_{i\alpha\downarrow}
  a_{j\beta\downarrow}a_{j\beta\uparrow}+\mbox{h.c.}\right)\right]
  \right\},
\end{equation}
\noindent
where the parameters are defined by the corresponding 
(primed) quantities in the atomic basis in the following 
manner
\begin{equation}
  \label{ea:dimm}
  \epsilon_a=\sum_{q\gamma}\left(\beta^{A\gamma}_q\right)^2\epsilon_a'+
  2\sum_{\stackrel{\scriptstyle q\gamma,r\delta}{q\gamma>r\delta}}
  \beta^{A\gamma}_q\beta^{A\delta}_{-r}{t'}^{\gamma\delta}_{q-r},
\end{equation}
\begin{equation}
  \label{t:dimm}
  t^{\alpha\beta}_p=\sum_{q\gamma}\beta^{\alpha\gamma}_q
  \beta^{\beta\gamma}_{p-q}\epsilon_a'+
  2\sum_{\stackrel{\scriptstyle q\gamma,r\delta}{q\gamma>r\delta}}
  \beta^{\alpha\gamma}_q\beta^{\beta\delta}_{p-r}
  {t'}^{\gamma\delta}_{q-r},
\end{equation}
$$
  U=\sum_{q\gamma}\left(\beta^{A\gamma}_q\right)^4U'+
  2\sum_{\stackrel{\scriptstyle q\gamma,r\delta}{q\gamma>r\delta}}
  \left[\left(\beta^{A\gamma}_q\beta^{A\delta}_r\right)^2\left(
  {K'}^{\gamma\delta}_{q-r}\right.\right.
$$
\begin{equation}
  \label{u:dimm}
  +2\left.\left.
  {J'}^{\gamma\delta}_{q-r}\right)+
  4\left(\beta^{A\gamma}_q\right)^3\beta^{A\delta}_r
  {V'}^{\gamma\delta}_{q-r}\right],
\end{equation}
$$
  K^{\alpha\beta}_p=\sum_{q\gamma}\left(
    \beta^{\alpha\gamma}_q\beta^{\beta\gamma}_{p-q}\right)^2U'+
  2\sum_{\stackrel{\scriptstyle q\gamma,r\delta}{q\gamma>r\delta}}
  \left\{\left(\beta^{\alpha\gamma}_q\beta^{\beta\delta}_{p-r}
  \right)^2{K'}^{\gamma\delta}_{q-r}\right.
$$
$$
  \left.
  +2\beta^{\alpha\gamma}_q\beta^{\beta\gamma}_{p-q}\left[
  \beta^{\alpha\delta}_r\beta^{\beta\delta}_{p-r}
  {J'}^{\gamma\delta}_{q-r}+
  \left(\beta^{\alpha\gamma}_q\beta^{\beta\delta}_{p-r}
  \right.\right.\right.
$$
\begin{equation}
  \label{k:dimm}
  +\left.\left.\left.
  \beta^{\beta\gamma}_{p-q}\beta^{\alpha\delta}_r\right)
  {V'}^{\gamma\delta}_{q-r}\right]\right\},
\end{equation}
$$
  V^{\alpha\beta}_p=\sum_{q\gamma}\left(
  \beta^{\alpha\gamma}_q\right)^3\beta^{\beta\gamma}_{p-q}U'+
  2\sum_{\stackrel{\scriptstyle q\gamma,r\delta}{q\gamma>r\delta}}
  \left\{\left(\beta^{\alpha\gamma}_q\right)^2\left[\beta^{\alpha\delta}_q
  \beta^{\beta\delta}_{p-r}
  \right.\right.
$$
$$
  \times\left.
  \left({K'}^{\gamma\delta}_{q-r}+
  {J'}^{\gamma\delta}_{q-r}\right)+
  \left(\beta^{\alpha\gamma}_q\beta^{\beta\delta}_{p-r}+
  3\beta^{\alpha\delta}_r\beta^{\beta\gamma}_{p-q}\right)
  {V'}^{\gamma\delta}_{q-r}\right]
$$
\begin{equation}
  \label{v:dimm}
  \left.
  +\beta^{\alpha\gamma}_q\left(\beta^{\alpha\delta}_r\right)^2
  \beta^{\beta\gamma}_{p-q}{J'}^{\gamma\delta}_{q-r}\right\},
\end{equation}
$$
  J^{\alpha\beta}_p=\sum_{q\gamma}\left(
  \beta^{\alpha\gamma}_q\beta^{\beta\gamma}_{p-q}\right)^2U'+
  2\sum_{\stackrel{\scriptstyle q\gamma,r\delta}{q\gamma>r\delta}}
  \left[\beta^{\alpha\gamma}_q\beta^{\alpha\delta}_r
  \beta^{\beta\gamma}_{p-q}\beta^{\beta\delta}_{p-r}
  \right.
$$
$$
  \times\left({K'}^{\gamma\delta}_{q-r}+{J'}^{\gamma\delta}_{q-r}\right)
  +\left(\beta^{\alpha\gamma}_q\beta^{\beta\delta}_{p-r}\right)^2
  {J'}^{\gamma\delta}_{q-r}
$$
\begin{equation}
  \label{j:dimm}
  +\left.
  2\beta^{\alpha\gamma}_q\beta^{\beta\gamma}_{p-q}\left(
  \beta^{\alpha\gamma}_q\beta^{\beta\delta}_{p-r}+
  \beta^{\alpha\delta}_r\beta^{\beta\gamma}_{p-q}\right)
  {V'}^{\gamma\delta}_{q-r}\right]. 
\end{equation}

\begin{figure}
\centering\epsfxsize=8cm\epsffile{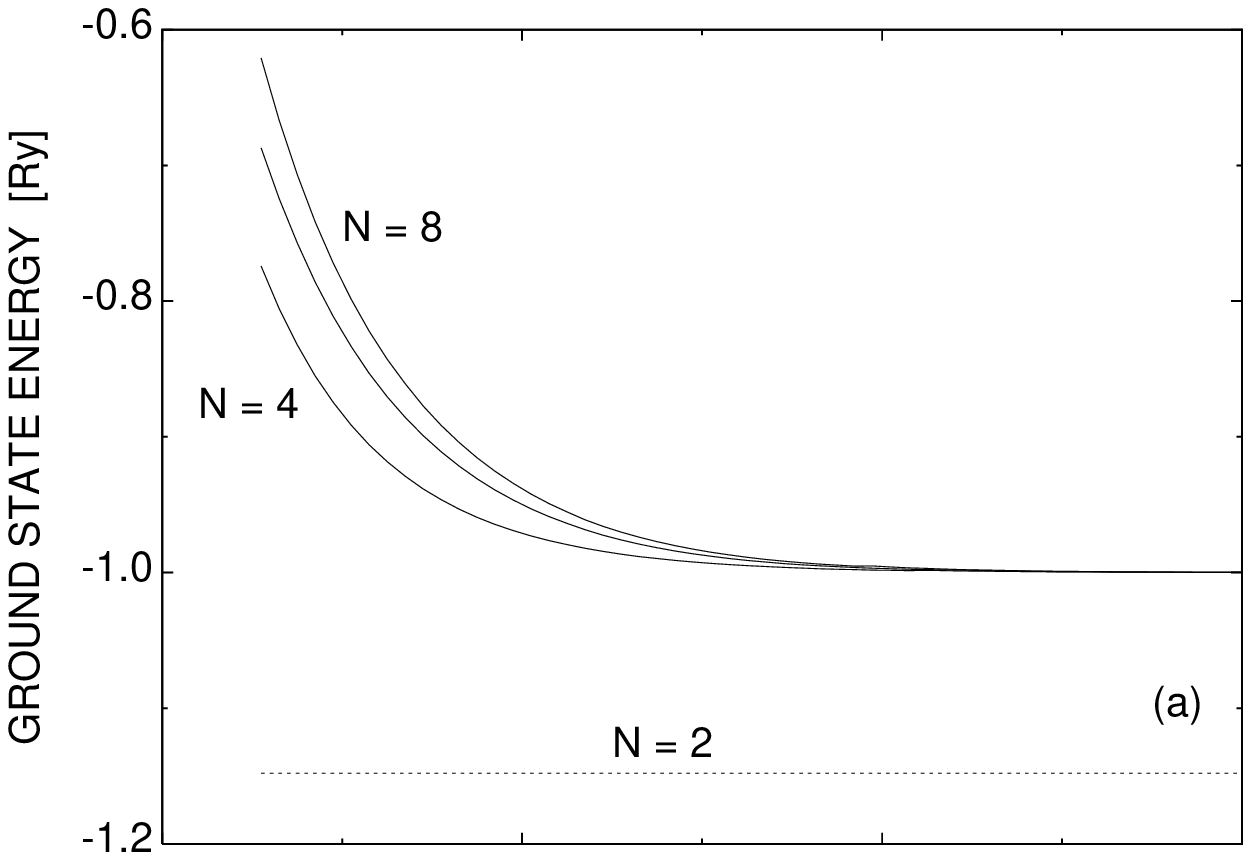} \propar
\centering\epsfxsize=8cm\epsffile{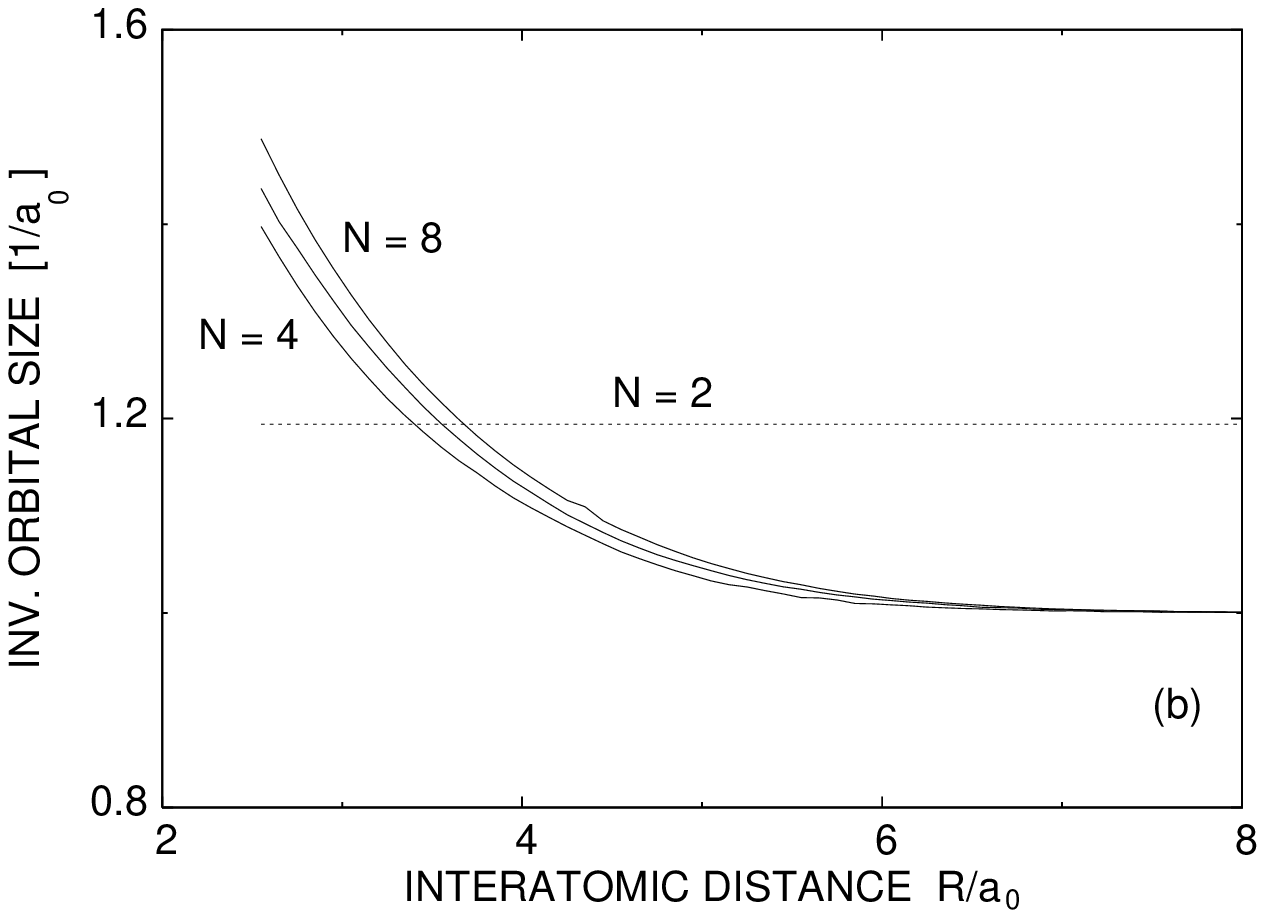} \par
\vspace{0.5cm}
\caption{
Ground state energy (a) and the optimal inverse orbital size (b)
of the dimerized chain versus $R$, 
for $N=2\div 8$ and for the optimized both the distorsion ($1-R_1/R$)
and the orbital size $\alpha^{-1}$. The dotted base lines mark 
the corresponding quantities for the $\mbox{H}_2$-molecule. 
\label{rys:egalpdimm}}
\end{figure}

Those microscopic parameters are calculated first in the 
atomic basis of $s$-type functions regarding the size $\alpha^{-1}$
of the orbitals {\em the same} on both sublattices. After the
calculations of the electronic ground-state 
energy have been finished, we include, as before, the
repulsion of (hydrogen) nuclei in the form (in atomic units):
\begin{equation}
  \label{jj:dimm}
  E_{N-N}\eqau
  \sum_{\stackrel{\scriptstyle i\alpha,j\beta}{i\alpha<j\beta}}
  \frac{2}{R^{\alpha\beta}_{ij}},
\end{equation}
where each nucleus has been taken into account only once. 
Adding the electronic and inter-nuclear parts, we compute
the ground state energy $E_G/N$ as a function of the average interatomic 
distance. In Fig.~\ref{rys:egalpdimm}a we show the ground-state energy for
$N=2$ (hydrogen molecule) and for the linear rings with even 
number of atoms ($N=4,6,8$); the discussion for odd number
of atoms requires 
a separate analysis, as the ground state configuration 
of nuclei is then a ionic-density wave 
with wave vector $Q<\pi/R$. This energy is minimized 
with respect to both $\alpha=\alpha_{\min}$ and the interatomic 
distance $R_1$ (cf.\ Fig.~\ref{rys:dimm}); 
in Fig.~\ref{rys:egalpdimm}b we display the inverse size (${\alpha}_{\min}$)
of the states (for $N=2-8$); they are quite close to those obtained in the 
undistorted case (i.e. now $\Delta\alpha<0.03$), so that the values
of the microscopic parameters such as $U$, $K_1$ or $t_1$ are not much 
different in both the distorted and the undistorted states. 
Also, our analysis, while confirming the existence of the Peierls
distorted state in the finite-size systems, shows that the distorsion fades
away with the increasing interatomic distance.

\subsection{Lattice dimerization vs.\ zero-point motion}

The zero-point motion of ions increases the system energy and
smears out their position. The dimerization diminishes
the system energy and, while shifting the ionic positions
relative to each other, it leaves their locations sharply 
defined. Both effects may be pronounced in the finite-size
systems, so the question arises how those two
effects compete with each other. 

\begin{figure}
\centering\epsfxsize=8cm\epsffile{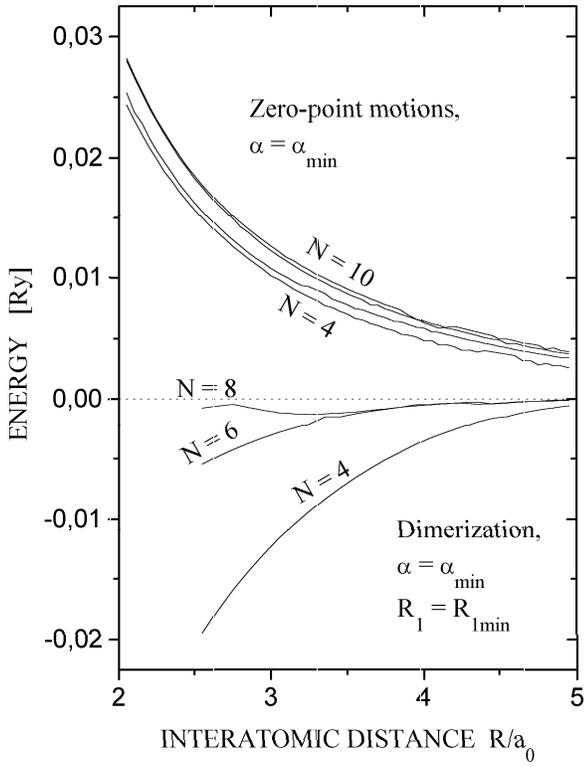} \par
\vspace{0.5cm}
\caption{
Ground-state-energy changes due to the dimerization
(bottom panel) and to the zero-point motion (top panel),
plotted for different $N$, as a function of $R$. 
\label{rys:deleg}}
\end{figure}

\begin{figure}
\centering\epsfxsize=8cm\epsffile{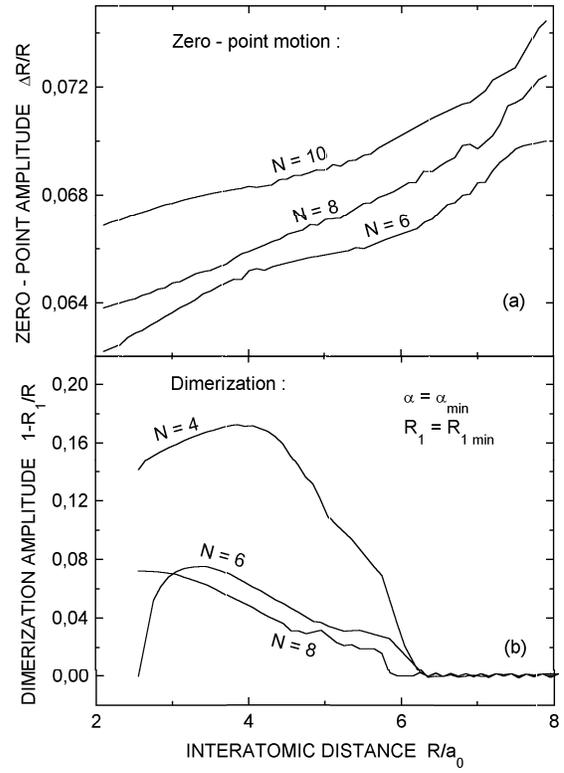} \par
\vspace{0.5cm}
\caption{
Atomic shift due to the dimerization and the 
root-mean-square amplitude of the zero point
motion vs.\ $R$ and for different $N$ 
(top and bottom parts, respectively).
\label{rys:delr}}
\end{figure}

In Fig.~\ref{rys:deleg} we compare the ground-state-energy changes
due to the dimerization and to the zero-point motion
for the atoms with hydrogen ionic mass. Whereas  the former 
decreases with the increasing size (bottom panel), the latter shows 
the opposite trend, as was discussed in the preceding Section. 
For comparison, the corresponding magnitudes of the average atomic shifts 
are displayed in Fig.~\ref{rys:delr}. Note that the
dimerization persists in the localized state of atoms and 
disappear only for $R/a_0>6$, when the Wannier functions
can be approximated with good accuracy by the atomic functions of
$1s$ type ($\alpha^{-1}\approx a_0$). The zero-point 
vibrations of the light atoms should also contribute to the blurring 
of picture of the charge densities observed e.g.\ in the scanning 
tunneling microscope of these nanoscopic objects.
This type of spectroscopy should be applied to the observation of those
effects on a local scale.

\section{A brief overview: New features}
\label{over}

In this work we have produced a fairly complete
description of one-dimensional model system by combining the {\em exact}
diagonalization of many-fermion Hamiltonian in the Fock
space with the subsequent {\em first-principle} readjustment of the
single-particle (Wannier) function. Electron and lattice
properties have been obtained 
as a function of the lattice parameter and the microscopic 
parameters have been determined explicitly. Our approach thus 
{\em extends} the current theoretical treatments 
\cite{hubb,poib,anis} to the {\em strongly correlated systems}
within the parametrized (second-quantized) models by providing the
determination of those parameters (coupling constants) and, in turn, 
determining the fundamental properties of the correlated 
state explicitly as a function of physical parameter, the lattice spacing $R$. 
Technically, we determine at each step 
the microscopic parameters taking the Wannier functions
with an adjusted size (starting from an exact Wannier functions for 
hydrogenic-like $s$-states), diagonalize the Hamiltonian in the Fock space 
with {\em all} pair interactions and {\em all} hopping 
integrals included, and thus obtained ground state energy is readjusted
again by changing variationally the size of the orbitals, 
calculating the changed parameters and performing 
again the diagonalization in the Fock space, and so on, until
the global minimum is reached for given interatomic 
distance (cf.\ Appendix~\ref{ground} for details). 
This procedure is then repeated for each selected interatomic distance. 

Our method of approach reveals features, 
which cannot be looked into when considering only the
parametrized models. Firstly, the atomic part of the energy
is not a constant, as it changes widely with the changing lattice
constant (cf.\ Table~\ref{tab:hopp}). 
Secondly, the crossover to the Mott localized 
state of electrons, as well as the lattice-dimerization 
evolution can be studied systematically, as the 
lattice expands. 
Thirdly, the effective masses for both band and
correlated states have been determined explicitly. 
Fourthly, and probably most importantly, 
the inclusion of {\em all} interactions (apart from the three- 
and the four-site terms) was possible, as all of them are 
calculated explicitly  (otherwise, the model would contain many 
parameters and become intractable or the results would be illegible).  

Such a procedure has been implemented so
far (on a desktop server) for one orbital-per-site model system
involving linear chains and rings containing up to $N=12$
atoms. However, due to the strong shrinking of
the Wannier functions induced by the strong correlations, 
the results are rapidly converging with $N$ and for $N\geq 8$ and
provide, in our view, a realistic estimate of the local 
properties of those strongly correlated systems. Obviously, 
we must incorporate the screening by
other than valence electrons, which are present in most of 
real systems, as well as to extend the method to 
$d$-type orbitals before the analysis will become
applicable to the correlated $3d$ systems at hand. Nonetheless, our 
analysis represents to the best of our knowledge, the 
first attempt to marrying consistently second- and first- and 
second-quantization aspects of the strong electronic correlations and as 
such should be tested in the clearest situation. 
Our method leads also to a renormalized wave equation \cite{spapo,ryspa}
for the single-particle wave function, but this feature of the method
requires a separate discussion.

The method is directly applicable to the correlated quantum dots, 
but here we avoided introducing the trapping potential, as
we would like to avoid mixing phenomenological and
microscopic concepts at this stage. 

In this paper we have considered only the
situation with one-electron per atom corresponding to the 
half-filled-band situation in the metallic state. Therefore, 
the onset of the Mott localization washes away any Luttinger-liquid
type of dynamics. Additionally, we have concentrated on 
static (equilibrium) properties by determining the ground-state
(spin-singlet) configuration and calculating its characteristics
such as statistical distribution function, spin-spin correlation
function, the exchange integrals, amplitude of dimerization, etc.
A direct determination of e.g.\ the Hubbard gap, quasiparticle 
properties, pairing tendencies or spin-spin charge 
separation requires calculation of the dynamic quantities such as the 
spectral-density function and the effective interaction, particularly 
for the systems with one or two electron holes in our starting system. 
The evolution of these properties as a function of interatomic
distance is of fundamental importance and
should be carried out next. 
Also, a detailed comparison of the behavior of systems with odd and even
number of electrons should be made. 

A separate question concerns the implementation of the 
wave-function optimization within the dynamical mean-field approach 
\cite{metz} in order to obtain 
an explicit mean-field solution of a $3d$ model system on the example 
of the $3d$ Hubbard model, as a function of the interatomic distance. 
This would also make possible 
a direct incorporation of the band-theoretical methods with those 
emphasizing the role of local electronic correlations. 
We should see a decisive progress in this matter soon.

Very recently, we have calculated the effective 
mass in the interacting system \cite{spary} which corresponds
to the quasiparticle mass in the infinite system. It is divergent
when the calculated distribution function $n_{k\sigma}$ is interpolated 
into a continuous distribution with a Fermi ridge. 
This results complement beautifully the results of
the approach presented in the present paper.

\section*{Acknowledgment}

The authors are grateful to their colleagues Dr.\ Robert Podsiad{\l}y 
and Dr.\ W{\l}odek W\'{o}jcik and Prof.\ K.\ Ro\'{s}ciszewski for many 
discussions. 
We are particularly grateful to Dr.\ Maciek Ma\'{s}ka from 
Silesian University for his insights on both analytical and
numerical aspects of the project. Dr.\ Marcello Acquarone 
acquainted us with various electron-lattice couplings. 
The comments made by Prof.\ Dieter Vollhardt from University 
of Ausburg and by Dr.\ Krzysztof Byczuk from Warsaw University 
are also appreciated. 
The work was supported by the State Committee for
Scientific Research (KBN) of Poland through Grant
No.\ 2P03B~092~18. One of us (A.R.) acknowledges 
the Estreicher Scholarship awarded to him by the 
Jagiellonian University. 

\appendix
\section{Microscopic parameters in the atomic basis}	
\label{micro}

In Ref.~\cite{ryspa} we have expressed the parameters 
$\epsilon_a$, $t_{ij}$, $U$, and $K_{ij}$ in the atomic basis
using the expansion \ref{wann}. Here we supplement them with the
formulae for the direct exchange integral $J_{i-j}$ and so-called
correlated hopping $V_{i-j}$. Namely, substracting \ref{wann} into
the expressions $J_p\equiv J_{i-j}=\bra{w_iw_j}V_{12}\ket{w_jw_i}$
and $V_p\equiv V_{i-j}=\bra{w_iw_i}V_{12}\ket{w_iw_j}$
we obtain recursively
$$
  J_p=\sum_q\beta_q^2\beta_{p-q}^2U'+
  4\sum_{\stackrel{\scriptstyle qr}{q>r}}
  \left(\beta_q\beta_{p-q}^2\beta_{r}\right.    
$$
$$
  +\left.
  \beta_q^2\beta_{p-q}\beta_{p-r}\right)V_{q-r}'
  +2\sum_{\stackrel{\scriptstyle qr}{q>r}}
  \beta_q\beta_{p-q}\beta_{p-r}\beta_r\left(J_{q-r}'\right.
$$
\begin{equation}
  \label{eatwan-lanext}
  +\left.
  K_{q-r}'\right)+2\sum_{\stackrel{\scriptstyle qr}{q>r}}
  \beta_q^2\beta_{p-r}^2J_{q-r}',
\end{equation}
$$
  V_p=\sum_q\beta_q^3\beta_{p-q}U'+
  2\sum_{\stackrel{\scriptstyle qr}{q>r}}\left(
  \beta_q^3\beta_{p-q}+3\beta_q^2\beta_{p-q}\beta_r\right)V_{q-r}'
$$
\begin{equation}
  +2\sum_{\stackrel{\scriptstyle qr}{q>r}}\beta_q^2\beta_r\beta_{p-r}
  \left(2J_{q-r}'+K_{q-r}'\right),
\end{equation}
\noindent
where $p$, $q$ and $r=0,1,\dots,N-1$, and the expansion 
coefficients $\beta_p$ of the Wannier functions obey the symmetry
$\beta_p=\beta_{-p}$ (thus for $N$ atoms there are $N/2$ or $(N-1)/2$ 
independent coefficients when $N$ is even or odd, respectively). 
The primed quantities are defined in the atomic basis 
$\left\{\Psi_j({\bf r})\right\}_{j=1,\dots,N}$. For example
$$
  J_{ij}'\equiv\bra{\Psi_i\Psi_j}V_{12}\ket{\Psi_j\Psi_i}=
  \int d^3{\bf r}d^3{\bf r}'\Psi_i^*({\bf r})\Psi_j^*({\bf r}')
$$
\begin{equation}
  \label{j:at}
  \times V_{12}({\bf r}-{\bf r}')\Psi_j({\bf r})\Psi_i({\bf r}').
\end{equation}
These expressions are used when evaluating the microscopic parameters
contained in the Hamiltonian. They are determined explicitly in the 
optimal state: $E_G=E_{\min}(R,\alpha=\alpha_{\min})$.

\section{Sigle-particle parameters in the atomic basis}
\label{single}

The expansion (\ref{wann}) leads to two-site terms in the atomic
energy $\epsilon_a$ and to and three-site terms in the hopping
integrals. For the sake of completness, we start from their
full expressions in the Wannier basis:
\begin{equation}
  \label{batwan-lanext}
  \epsilon_a=\sum_{q}\beta_q^2\epsilon_a'+
  2\sum_{\stackrel{\scriptstyle qr}{q>r}}\beta_q\beta_rt_{q-r}',
\end{equation}
\begin{equation}
  t_p=\sum_{q}\beta_q\beta_{p-q}\epsilon_a'+
  2\sum_{\stackrel{\scriptstyle qr}{q>r}}\beta_q\beta_{p-r}t_{q-r}'.
\end{equation}
In these expressions the atomic energy $\epsilon_a'$ 
(in atomic units) is
$$
  \epsilon'_a=\left<\Psi_i\left|-\nabla^2
  -\sum_j\frac{2}{r_j}\right|\Psi_i\right>=
$$
\begin{equation}
  \label{eapgen}
  =\alpha^2-2\alpha+\sum_{j\neq i}\left[-\frac{2}{R_{ij}}
  +\exp(-2\alpha R_{ij})\left(2\alpha+\frac{2}{R_{ij}}\right)\right],
\end{equation}
\noindent
where $R_{ij}=\left|{\bf R}_i-{\bf R}_j\right|$
and ${\bf r}_j=\left|{\bf r}-{\bf R}_j\right|$. 
Note the appearance of the long-range 
part $\sim\left(-2/R_{ij}\right)$, as one would have in the classical 
limit. 

\begin{figure}
\centering\epsfxsize=8cm\epsffile{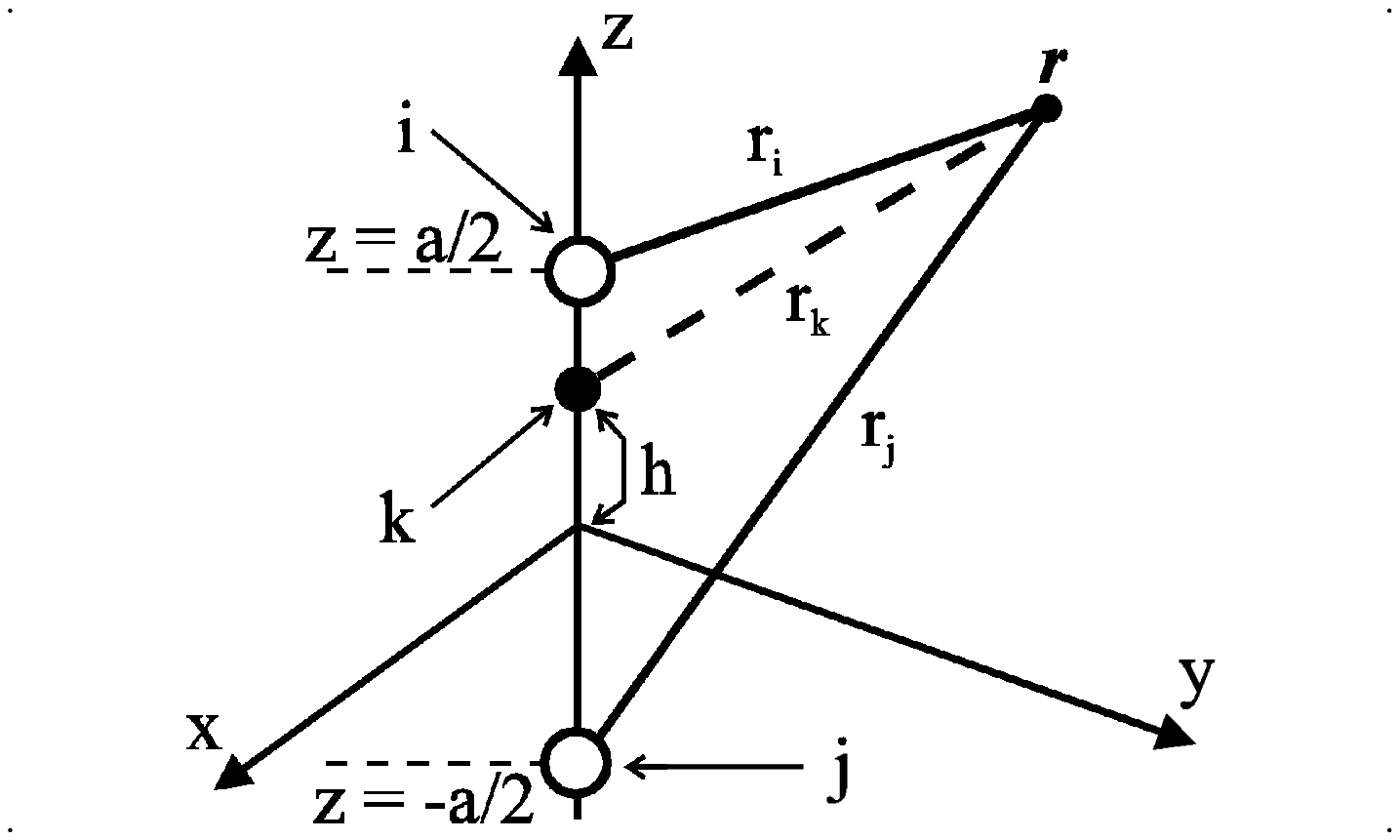} \par
\vspace{0.5cm}
\caption{
Configuration of coordinates used to 
calculated the three- site terms $\tau_{ikj}$ in the hopping
integral $t_{ij}'$ for the electron transfer $j\rightarrow i$ induced
by the Coulomb potential of $k$-th ion.
\label{rys:tau}}
\end{figure}

The evaluation of $t_{ij}'$ is not so straightforward 
as one can see from the expression
\begin{equation}
  \label{tpgen}
  t'_{ij}=\left<\Psi_i\left|-\nabla^2
  -\sum_k\frac{2}{r_k}\right|\Psi_j\right>=\tau_0-2\sum_k \tau_{ikj},
\end{equation}
where $r_k\equiv\left|{\bf r}-{\bf R}_k\right|$, 
$\tau_0$ represent the simple
part and $\tau_{ijk}$ is the three-site part. 
The part $\tau_0$ is easy to calculate, since
\begin{equation}
  \label{tauo}
  \tau_0\equiv \left<\Psi_i\right|-\nabla^2\left|\Psi_j\right>
  =\alpha^2e^{-\alpha R_{ij}}\left(1+\alpha R_{ij}
  -\frac{1}{3}\alpha^2R_{ij}^2\right).
\end{equation}
The three-site part is more cumbersome, as it reduces to 
the following integral expression
\begin{equation}
  \label{tauikj}
  \tau_{ikj}\equiv\int d^3r\Psi^*({\bf r}_i)\frac{1}{r_k}\Psi({\bf r}_j)=
  \frac{\alpha^3}{\pi}\int d^3r\frac{e^{-\alpha(r_i+r_j)}}{r_k}.
\end{equation}
To calculate the integral we introduce the
spheroidal coordinates $(\lambda,\mu)$
$$
  a\lambda=r_i+r_j,\ \ \ \  
  a\mu=r_i-r_j,
$$
\begin{equation}
  \label{sferlmu}
  d^3r=\frac{\pi a^3}{4}(\lambda^2-\mu^2)d\lambda d\mu,
\end{equation}
where $a\equiv R_{ij}$. The regimes for $\lambda$ and $\mu$ are:
$1<\lambda<\infty$, $-1<\mu<1$. 
This transformation leads to the following expression for $r_k$
\begin{equation}
  \label{sferrk}
  r_k=\sqrt{\frac{a^2}{4}(\lambda^2-1)(1-\mu^2)
  +\left(\lambda\mu\frac{a}{2}-h\right)},
\end{equation}
where $h$ is the $z$-coordinate of the middle point of the 
Coulomb potential well
caused by the $k$-th ion (cf.\ Fig~\ref{single}\ref{rys:tau}). 
Integrating with respect to $\mu$ we obtain
$$
  \tau_{ikj}=\frac{1}{2}\alpha^3a^2\int_1^{\infty}d\lambda 
  e^{-\alpha a\lambda}\left\{\left[\lambda^2\left(1+(2h/a)^2\right)
  +b/2\right]\times\right.
$$
$$
  \times\log\frac{(2h\lambda/a-1)-\sqrt{(2h\lambda/a-1)^2+b}}{(2h\lambda/a+1)
  -\sqrt{(2h\lambda/a+1)^2+b}}+
$$
$$
  +(3h\lambda/a-1/2)\sqrt{(2h\lambda/a+1)^2+b}
$$
\begin{equation}
  \label{sfertau}
  \left.
  -(3h\lambda/a+1/2)\sqrt{(2h\lambda/a-1)^2+b}\right\},
\end{equation}
\noindent
where $b=(\lambda^2-1)\left[1-(2h/a)^2\right]$. This integral simplifies
substantially if either $k=i$ or $k=j$, i.e.\ for $h=\pm a/2$. 
Then, taking simple limiting expression we obtain that
\begin{equation}
  \label{diagtau}
  \tau_{ijk}=\alpha^3a^2\int_1^{\infty}d\lambda\lambda
  e^{-\alpha a\lambda}=\alpha e^{-\alpha R_{ij}}(\alpha R_{ij}+1).
\end{equation}
Substituting (\ref{diagtau}) to (\ref{tpgen}) we obtain the expression for the
hopping integral for ${\rm H}_2$ molecule and for the linear chain in 
the tight-binding approximation (cf.\ Ref.~\cite{spapo}).

In the general case, we have to evaluate the integral 
(\ref{sfertau}) numerically. For this purpose, one makes the change
of variable $\lambda=1/t$ to integrate over the finite interval
$0<t<1$ (one has to use a variable summation step, since 
the integrand is logarithmically divergent at $t=1$). 
When using the Simpson method \cite{burd} one
has to evaluate the integrand in $300\div 400$ points to achieve
the accurancy $10^{-6}\mbox{ Ry}$; this procedure requires a neglegible 
time compared to that required to determine
the ground state energy. 

Numerical calculations of {\em all} hoppings 
$\left\{t_{ij}\right\}_{i,j=1,\dots,N}$ 
can be accelerated by exploiting the problem 
symmetry. Among $N(N-1)/2$ possible hoppings for
the linear-chain case with periodic boundary conditions 
($t_{ij}=t_{i-j}$), we have only $(N-N\bmod 2/N)$ different
integrals. In case of dimerization this number is doubled, 
as the hopping elements change 
$t_{ij}\rightarrow t_{ij}^{AA}=t_{ij}^{BB},t_{ij}^{AB}=t_{ji}^{BA}$. 
Additionally, we use the symmetry $h\rightarrow -h$ when evaluating 
$\tau_{ijk}$.

\section{Ground-state energy evaluation in the Fock space}
\label{ground}

For $N$ lattice sites and the grand canonical system with 
variable number of electrons $N_e=0,1,\dots,2N$ and the total number
of electrons with the spins up $N_{\uparrow}=0,1,\dots,N_e$, we have
the Fock (occupation-number) space  ${\cal H}_1\equiv {\cal H}_1
\{N,N_e=0,\dots,2N,N_{\uparrow}=0,\dots,N_e\}$
of dimension $4^N$. Additionally, if the number of
electrons is fixed at $N_e=N$ (i.e.\ for the case of one
electron per atom) then the Fock space ${\cal H}_2\equiv 
{\cal H}_2\{N,N_e=N,N_{\uparrow}=0,\dots,N_e\}$ has reduced dimension to
$\dim {\cal H}_2=\newton{2N}{N_e}$.
Moreover, if the total
$z$-component of spin is $S_{\rm tot}^z=0$ (for $N$ even, as happens
in the situation then the space ${\cal H}_2\{\dots\}$) reduces
to the space ${\cal H}_3\{N,N_e=N,N_{\uparrow}=N/2\}$, with
the corresponding dimension $\dim {\cal H}_3=\newton{N}{N/2}^2$. Those
dimensions for $N=10$ are, respectively: 
$\dim {\cal H}_1=1\ 048\ 567$, $\dim {\cal H}_2=184\ 756$, and
$\dim {\cal H}_3=63\ 504$. The implementation of the translational 
symmetry does not influence essentially either the
computing time or the memory capacity required. 

The basis vectors in the Fock space are 
represented by the site (Wannier-state) occupancies 
$\left\{n_{i\sigma}\right\}_{\stackrel{\scriptstyle i=1,\cdots,N}%
{\sigma=\uparrow,\downarrow}}$ as
\begin{equation}
  \label{fockvec}
  \left|v\right>=\left|n_{1\uparrow},...,n_{N\uparrow},
  n_{1\downarrow},...,n_{N\downarrow}\right>.
\end{equation}
The creation and annihilation operators are defined 
in a standard manner:
\begin{equation}
  \label{kredef}
  a^{\dagger}_{i\sigma}\left|...,n_{i\sigma},...\right>=
  (-1)^{\displaystyle\nu_{i\sigma}}\left(1-n_{i\sigma}\right)
  \left|...,n_{i\sigma}+1,...\right>,
\end{equation}
\begin{equation}
  \label{anidef}
  a_{i\sigma}\left|...,n_{i\sigma},...\right>=
  (-1)^{\displaystyle\nu_{i\sigma}}n_{i\sigma}
  \left|...,n_{i\sigma}-1,...\right>,
\end{equation}
where $\nu_{i\sigma}$ is the number of electrons in the states preceding
the $i$-th site (including
the opposite spins if $\sigma=\downarrow$). We have also used the notation
$0\ket{\dots}\equiv 0$. The above definitions allow for an unambigous
determination of the matrix representation of the Hamiltonian. Simply,
we have to determine the matrix elements $\bra{u}{\cal H}\ket{v}$
for $u,v\in{\cal H}_3$. In practice, the basis vectors 
$\left\{\ket{v}\right\}$
are ordered as the combination series. Therefore, acting on each of
them with consecutive terms of ${\cal H}$, we
immediatly can identify the result of the action 
as proper basis vector $\left\{\ket{u}\right\}$ (or zero vector). Thus, the
number of operations is equal to the number of nonzero elements 
$\left\{\bra{u}{\cal H}\ket{v}\right\}$. 

The method allows for the construction of the Hamiltonian 
matrix representation for which the lowest 
eigenvalue is the ground state energy. The physics 
of the problem is determined usually by
$\sim 10\%$ of the eigenvectors with the lowest eigenenergies. 
In this situation the Lanczos method \cite{nish} in the
version proposed by Nishino \cite{nish} is most appropriate. 
In its latter version one uses the {\em cummulant expansion} 
containing up to $10^2\div 10^3$ terms for $N=6\div 10$
atoms with the lattice constant $R=2\div 3 a_0$ (for larger
$R$ and/or smaller $N$ the number of terms to be included 
is much smaller). The numerical accurancy of the result for
$E_G$ was typically $10^{-6}\mbox{ Ry}$. 

The final step involved the minimization of $E_G$
with respect to $\alpha$ and (whenever possible) with respect to $R$ 
(in the case of dimerization it is the shorter lattice constant $R_1$). 
The minimization was carried out using respectively one- or 
two-dimensional {\em simplex method} \cite{bran}, with accurancy of the
order $10^{-5}\mbox{ Ry}$. In practice, to achieve such accuracy 
in the minimization process with respect to $\alpha$ (and $R_1$) one has
to repeat the whole procedure $20\div 40$ times for the normal state 
and $100\div 200$ times for the dimerized state. The non-dimerized 
state has no minimum with respect to $R$; this problem is elaborated
on in main text (see also \cite{spapo,ryspa}). 

One more remark. To utilize the discrete translational 
symmetry by shifting state $\ket{u}$ by one lattice constant 
we have to introduce the basis of eigenstates of the corresponding operator
${\cal R}$, which are defined by
\begin{equation}
  \label{baseuk}
  \left|u; k\right>\equiv\frac{1}{N^{1/2}}\sum_{j=0}^{N-1}
  \exp\left(i\frac{2\pi kj}{N}\right){\cal R}^j\left|u\right>,
\end{equation}
corresponding to the eigenvalues $\exp (i2\pi k/N)$ labeled
by $k=0,1,\dots,N-1$. Under such circumstances, we can
consider the construction of the Hamiltonian matrix in subspaces
of dimension $\dim {\cal H}_3/N$ each. But then, we have to perform
additional summation over $j$, so the number of nonzero
elements will increase by factor $\sim N$. In effect, we do not
gain much. However, in the planned by us calculations of the excited
states this reduction of the Hamiltonian dimension is crucial
in making the diagonalization method effective even thought
the number of nonzero elements remains the same. 
Here we do not make use of the translational symmetry.


\newpage
\begin{center}
{\bf FIGURES}
\end{center}

\begin{figure}
\centering\epsfxsize=8cm\epsffile{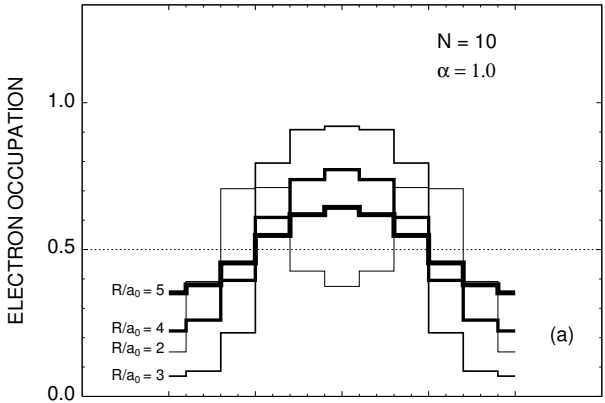} \propar
\centering\epsfxsize=8cm\epsffile{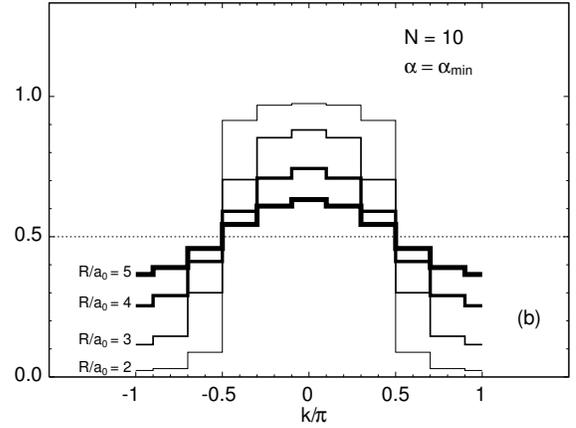} \par
\vspace{0.5cm}
\caption{
The statistical distribution function $n_{k\sigma}$ for $N=10$ 
in the state with optimized orbitals (a) and with 
non-optimized orbitals (b); both for different interatomic 
distances $R$.
\label{rys:nks}}
\end{figure}

\vspace{2in}

\newpage

\end{multicols}


\begin{thebibliography}{99}
\bibitem{jogu}
R.~O.~Jones and O.~Gunnarson, 
Rev.\ Mod.\ Phys.\ {\bf 61}, 689 (1989);
W.~Pickett, {\em ibid}.\ {\bf 61}, 433 (1989); 
W.~M.~Temmerman, A.~Svane, Z.~Szostek and H.~Winter, 
in {\em Electronic Density Functional Theory: 
Recent Progress and New Directions}, 
edited by J.~F.~Dobson, G.~Vignale, and H.~Das 
(Plenum, New York 1998) pp. 327-47.

\bibitem{hubb}
The question of importance of intersite interaction 
within the context 
of parametrized models for low-dimensional systems
was addressed by a number of authors, see:
J.~Hubbard, Phys.\ Rev.\ B {\bf 17}, 494 (1978);
J.~Kondo and K.~Yamai, 
J.\ Phys.\ Soc.\ Japan {\bf 43}, 424 (1977);
A.~A.~Ovchinnikov, 
Mod.\ Phys.\ Lett.\ B, {\bf 7} 1397 (1993);
S.~Caponi, D.~Poilblanc, and T.~Giamarchi, 
Phys.\ Rev.\ B {\bf 61}, 13410 (2000);
R.~Strack and D.~Vollhardt, 
Phys.\ Rev.\ Lett.\ {\bf 70}, 2637 (1993);
L.~Arachea and A.~A.~Aligia, 
Phys.\ Rev.\ Lett.\ {\bf 73}, 2240 (1994).

\bibitem{schu}
For review see: 
H.~J.~Schultz, 
in {\em Correlated Electron Systems}, 
edited by V.~J.~Emery 
(World Scientific, Singapore, 1993) p.\ 199ff;
J.~Voit, Rep.\ Prog.\ Phys.\ {\bf 57}, 977 (1995).
Experimentally: 
C.~Kim, Z.-X.~Shen, N.~Motoyama, H.~Eisaki, S.~Uchida, 
T.~Tohyama and S.~Maekawa, 
Phys.\ Rev.\ B {\bf 56}, 15589 (1997).

\bibitem{poib}
D.~Poilblanc, S.~Yunoki, S.~Maekawa, and E.~Dagotto, 
Phys.\ Rev.\ B {\bf 56}, R1645 (1997); 
T.~Giamarchi and A.~J.~Millis,
Phys.\ Rev.\ B {\bf 46}, 9325 (1992);
S.~Daul and R.~M.~Noack,
Phys.\ Rev.\ B {\bf 61} 1646 (2000). 

\bibitem{peie}
R.~E.~Peierls, 
{\em Quantum Theory of Solids}
(Claredon Press, Oxford, 1953) p.\ 108ff;
M.~J.~Rice and S.~Str\"assler, 
Solid State Commun. {\bf 13}, 125 (1973);
for the discusion of relation 
to the charge- and spin-density wave states see e.g. 
S.~Caprara, M.~Avignon, and O.~Navarro, 
Phys.\ Rev.\ B {\bf 61}, 15667 (2000).
The quantum fluctuation in noncorrelated systems were discussed in:
J.~E.~Hirsh and E.~Fradkin, 
Phys.\ Rev.\ Lett.\ {\bf 49}, 402 (1982).

\bibitem{goni}
A.~R.~Goni, A.~Pinczuk, J.~S.~Weiner, B.~S.~Dennis, 
L.~N.~Pfeifer, and K.~West, 
Phys.\ Rev.\ Lett.\ {\bf 70}, 1151 (1993);
for review see:
T.~Ishiguro, K.~Yamaji, and G.~Saito, 
{\em Organic Superconductors} 
(Springer Verlag, Berlin, 1998).

\bibitem{mint}
J.~W.~Mintmire, B.~I.~Dunlap, and C.~T.~White,
Phys.\ Rev.\ Lett.\ {\bf 68}, 631 (1992).

\bibitem{spapo}
J.~Spa{\l}ek, R.~Podsiad{\l}y, W.~W\'{o}jcik, and A.~Rycerz,
Phys.\ Rev.\ B {\bf 61}, 15676 (2000);
for a brief review see:
J.~Spa{\l}ek {\em et al.}, 
Acta Phys.\ Polonica B {\bf 31}, 2879 (2000).

\bibitem{ryspa}
A.~Rycerz and J.~Spa{\l}ek, 
Phys.\ Rev.\ B {\bf 63}, 073101 (2001).

\bibitem{mott}
N.~F.~Mott, 
{\em Metal-Insulator Transitions}
(Taylor and Francis, London, 1990).

\bibitem{spawo}
See e.g. 
J.~Spa{\l}ek, J.\ Solid St.\ Chem.\ {\bf 88}, 70 (1990);
J.~Spa{\l}ek and W.~W\'{o}jcik, in 
{\em Spectroscopy of Mott Insulators and Correlated Metals},
edited by A.~Fujimori and Y.~Tokura
(Springer Verlag, Berlin, 1995) p.\ 41ff.

\bibitem{migd}
The existence of the Fermi level 
appears also for finite $N$ as it expresses 
the Pauli principle for quasiparticles 
(exists for example for nuclei).
The existence of the Fermi discontinuity in interacting gas
has been discussed by 
A.~B.~Migdal, 
J.\ Exp.\ Teoret.\ Phys.\ (USSR) {\bf 32}, 399 (1957)
[Sov.\ Phys.\ -~JETP {\bf 5}, 333 (1957)]; 
E.~Daniel and S.~Vosko, Phys.\ Rev.\ {\bf 120}, 2041 (1960);
J.~M.~Luttinger, Phys.\ Rev.\ {119} 1153 (1960);
for an interpolation between the high- and 
the low-density limits see e.g.
N.~H.~March, W.~H.~Young, and S.~Sampantar, 
{\em The Many-Body Problem in Quantum Mechanics}
(Dover, New York, 1995) pp.\ 166-8.

\bibitem{ande}
The kinetic exchange integrals defined in this manner 
are well defined only in the limit 
$\left|t_m+V_m\right|\ll U-K_m$ 
For $m=1$ and neglecting $K_m$ they were defined in
P.~W.~Anderson, Phys.\ Rev.\ {\bf 115}, 2 (1959).
In general case, they were considered in:
J.~Spa{\l}ek, A.~M.~Ole\'{s}, and K.~A.~Chao,
phys.\ stat.\ solidi (b) {\bf 108}, 329 (1981).

\bibitem{birg}
C.f.\ R.~J.~Birgenau and G.~Shirane in 
{\em Physical Properties of High-Temperature Superconductors},
edited by D.~M.~Ginsberg
(World Scientific, Singapore, 1989) vol.~1, p.\ 151ff
and references therein.

\bibitem{ovch}
The discussion here systematizes to some extent various types 
of local electron-lattice couplings considered earlier:
A.~A.~Ovchinnikov, Fiz.\ Tverd.\ Tela {\bf 1}, 832 (1965)
[Sov.\ Phys.\ -~Solid State {\bf 7}, 664 (1965)];
T.~Holstein, Ann.\ of Phys.\ {\bf 8}, 325 (1959);
S.~Bari\v{s}i\'{c}, Phys.\ Rev.\ B {\bf 22}, 2099 (1980);
M.~Acquarone and C.~Noce, 
Int.\ J.\ Mod.\ Phys.\ B {\bf 13}, 3331 (1999).

\bibitem{nish}
T.~Nishino and J.~Kanamori, 
J.\ Phys.\ Soc.\ Japan {\bf 59}, 253 (1990);
T.~Nishino, 
{\em Electron Correlation Effects in Low Dimensional
Periodic Systems}, unpublished. 
It is based on the Lanczos method: 
C.~Lanczos, 
J.\ Res.\ Nat.\ Bur.\ Std.\ {\bf 45}, 255 (1950);
see also:
J.~K.~Cullum and R.~A.~Willonghby, 
{\em Lanczos Algorithms for Large Symmetric Eigenvalue 
Computation} (Birkhauser, Boston, 1985);
E.~Dagotto and A.~Moreo, Phys.\ Rev.\ D {\bf 31}, 865 (1985).

\bibitem{bran}
S.~Brandt, 
{\em Statistical and Computational Methods in Data Analysis}
(Springer Verlag, New York, 1997).

\bibitem{burd}
See, e.g. 
R.~L.~Burden, J.~D.~Faires, {\em Numerical Analysis} 
(Prindle, Weber \& Schmidt, Boston, 1985).

\bibitem{anis}
Earlier efforts contain an approximate treatment
of correlations in $3d$ systems combined with
the ab initio calculations of single-particle states; see:
V.~I.~Anisimov, J.~Zaanen, and O.~K.~Andersen, 
Phys.\ Rev.\ B {\bf 44}, 943 (1991) 
-~the so-called LDA+U; 
J.~Spa{\l}ek and W.~W\'{o}jcik, 
Phys.\ Rev.\ B {\bf 45}, 3799 (1992)
-~Gutzwiller approach to correlations;
G.~Stollhoff, Europhys.\ Lett.\ {\bf 30}, 99 (1995) 
-~the so-called local-ansatz+LDA. 

\bibitem{metz}
W.~Metzner and D.~Vollhardt, 
Phys.\ Rev.\ Lett.\ {\bf 62}, 324 (1989);
M.~Jarrell, {\em ibid}.\ {\bf 68}, 168 (1992). 

\bibitem{spary}
J.~Spa{\l}ek and A.~Rycerz, submitted to Phys.\ Rev.\ Lett.
\end{thebibliography}
\end{document}